\documentclass[useAMS,usenatbib]{mn2e}
\usepackage[dvips]{graphicx}
\usepackage{amssymb}
\usepackage{txfonts}

\newcommand{\ft}{FORS2}

\newcommand{\xs}{X-Shooter}
\newcommand{\sw}{{\it Swift}}
\newcommand{\ob}{{\it o}}
\newcommand{\eb}{{\it e}}

\title[Polarimetry of the afterglow of GRB 091018]{Detailed optical and near-infrared polarimetry, spectroscopy and 
broadband photometry of the afterglow of GRB 091018: Polarisation evolution}

\author[K.~Wiersema et al.]
{\parbox{\textwidth}{K. Wiersema,$^{1}$\thanks{E-mail: kw113@star.le.ac.uk}
P. A. Curran,$^{2}$
T. Kr\"{u}hler,$^{3}$
A. Melandri,$^{4,5}$ 
E. Rol,$^{6}$
R. L. C. Starling,$^{1}$
N. R. Tanvir,$^{1}$
A. J. van der Horst,$^{6}$
S. Covino,$^{4}$
J. P. U. Fynbo,$^{3}$
P. Goldoni,$^{7}$
J. Gorosabel,$^{8}$
J. Hjorth,$^{3}$
S. Klose,$^{9}$
C. G. Mundell,$^{5}$
P. T. O'Brien,$^{1}$
E. Palazzi,$^{10}$
R. A. M. J. Wijers,$^{6}$
V. D'Elia,$^{11,12}$
P. A. Evans,$^{1}$
R. Filgas,$^{13}$
A. Gomboc,$^{14,15}$
J. Greiner,$^{13}$
C. Guidorzi,$^{16}$
L. Kaper,$^{6}$
S. Kobayashi,$^{5}$
C. Kouveliotou,$^{17}$
A. J. Levan,$^{18}$
A. Rossi,$^{9}$
A. Rowlinson,$^{1,6}$
I. A. Steele,$^{5}$
A. de Ugarte Postigo$^{3,8}$ and
S. D. Vergani$^{4}$
}\vspace{0.4cm}\\
\parbox{\textwidth}{$^{1}$ University of Leicester, University Road, Leicester LE1 7RH, UK\\
$^{2}$ Laboratoire AIM, CEA/IRFU-Universit\'e Paris Diderot-CNRS/INSU, CEA DSM/IRFU/SAp, Centre de Saclay, F-91191 Gif-sur-Yvette, France\\
$^{3}$ Dark Cosmology Centre, Niels Bohr Institute, University of Copenhagen, Juliane Maries Vej 30, 2100 Copenhagen, Denmark\\
$^{4}$ INAF - Osservatorio Astronomico di Brera, via E. Bianchi 46, I-23807 Merate, Italy\\
$^{5}$ Astrophysics Research Institute, Liverpool John Moores University, Twelve Quays House, Egerton Wharf, Birkenhead CH41 1LD, UK\\
$^{6}$ Astronomical Institute $''$Anton Pannekoek$''$, University of Amsterdam, 1090 GE Amsterdam, The Netherlands\\
$^{7}$ Laboratoire Astroparticule et Cosmologie, 10 rue A. Domon et L. Duquet, 75205 Paris Cedex 13, France\\
$^{8}$ IAA - CSIC, Glorieta de la Astronom\'{i}a s/n, 18008 Granada, Spain\\
$^{9}$ Th\"{u}ringer Landessternwarte Tautenburg, Sternwarte 5, 07778 Tautenburg, Germany\\
$^{10}$ INAF - Istituto di Astrofisica Spaziale e Fisica Cosmica di Bologna, Via Gobetti 101, I-40129 Bologna, Italy\\
$^{11}$ INAF - Osservatorio Astronomico di Roma, via di Frascati 33, 00040 Monte Porzio Catone, Italy\\
$^{12}$ ASI-Science Data Center, via Galileo Galilei, 00044 Frascati, Italy\\ 
$^{13}$ Max-Planck-Institut f\"{u}r Extraterrestrische Physik, Giessenbachstra{\ss}e 1, 85748 Garching, Germany\\
$^{14}$ Faculty of Mathematics and Physics, University of Ljubljana, Jadranska 19, SI-1000 Ljubljana, Slovenia\\
$^{15}$ Centre of Excellence SPACE-SI, A\v{s}ker\v{c}eva cesta 12, SI-1000 Ljubljana, Slovenia\\
$^{16}$ Physics Department, University of Ferrara, via Saragat 1, I-44122, Ferrara, Italy \\
$^{17}$ Space Science Office, VP62, NASA/Marshall Space Flight Center Huntsville, AL 35812, USA\\
$^{18}$ Department of Physics, University of Warwick, Coventry CV4 7AL, UK
 }}

\begin{document}

\date{Submitted xx/xx/2012}

\pagerange{\pageref{firstpage}--\pageref{lastpage}} \pubyear{2012}

\maketitle

\label{firstpage}

\begin{abstract}
Follow-up observations of large numbers of gamma-ray burst (GRB) afterglows, facilitated by the \sw\ satellite, have produced a large sample of spectral energy distributions and light curves, from which their basic micro- and macro-physical parameters can in principle be derived.  However, a number of phenomena have been observed that defy explanation by simple versions of the standard fireball model, leading to a variety of new models.
Polarimetry can be a major independent diagnostic of afterglow physics, probing the magnetic field properties and internal structure of the GRB jets.
In this paper we present the first high quality multi-night polarimetric light curve of a \sw\ GRB afterglow, aimed at providing a well calibrated dataset of a typical afterglow to serve as a   
benchmark system for modelling afterglow polarisation behaviour.  
In particular, our dataset of the afterglow of GRB\,091018 (at redshift $z=0.971$) comprises optical linear polarimetry ($R$ band, $0.13 - 2.3$ days after burst); circular polarimetry ($R$ band) and near-infrared linear polarimetry ($Ks$ band). We add to that high quality optical and near-infrared broadband light curves and spectral energy distributions as well as afterglow spectroscopy.
The linear polarisation varies between 0 and 3\%, with both long and short time scale variability visible.
We find an achromatic break in the afterglow light curve, which corresponds to features in the polarimetric curve. We find that the data can be reproduced by jet break models only if an additional polarised component of unknown nature is present in the polarimetric curve. 
We probe the ordered magnetic field component in the afterglow through our deep circular polarimetry, finding $P_{\rm circ} < 0.15$\% (2$\sigma$), the deepest limit yet for a GRB afterglow, suggesting ordered fields are weak, if at all present. Our simultaneous $R$ and $Ks$ band polarimetry shows that dust induced polarisation in the host galaxy is likely negligible.
\end{abstract}


\section{Introduction \label{sec:intro}}
Collimated outflows in the form of jets are ubiquitous, from AGN jets driven by accretion of material by supermassive black holes in galactic centres to those associated with stellar sources such as X-ray binaries and galactic microquasars. A particularly exciting view of the fundamental physics of jet sources is offered by gamma-ray bursts (GRBs), which form the extreme end of the energy- and Lorentz factor parameter space.
Since the discovery of afterglows in 1997 (\citealp{vanparadijs,costa}) we have a broad picture of the origin and cause of GRBs: through a catastrophic event (in the case of long-duration GRBs this is the core-collapse of a massive star), a jet of highly relativistic material is ejected. 
In the standard fireball model (e.g. M\'{e}sz\'{a}ros \& Rees 1997), the resulting broadband emission detected from X-ray energies through to radio waves - the so-called afterglow - is explained by the relativistic ejecta colliding with the surrounding circumburst medium. 
The afterglow radiation we detect is consistent with a synchrotron emission  (e.g. M\'{e}sz\'{a}ros \& Rees 1992, 1997; Wijers \& Galama 1999), characterised by a series of smoothly connected power laws, with characteristic break frequencies and fluxes. 
The macroscopic properties of shocks are largely understood, and the dynamics of the shock created when the 
relativistic jet hits the circumstellar matter can be written down in terms of the explosion energy, the density (and density gradient) of 
the medium into which the shock ploughs, and the composition of the shocked material. 
However, oustanding questions remain on the nature of the \emph{micro}physics: how are the relativistic particles that 
radiate the observed emission accelerated? Where does the magnetic field in the shocked region 
come from and what is its structure?

Since its launch in 2004, the {\em Swift} satellite (\citealp{Gehrels}) has provided us with hundreds of well sampled X-ray and UV/optical afterglow light curves from seconds to
months after burst (see for a review \citealp{Gehrelsreview}). These light curves revealed complexity that was unexpected from pre-{\em Swift} light curves (\citealp{Piranreview}); the standard fireball model has received a growing number of modifications, and the concept of a canonical light curve was introduced to explain the steep fades, plateau phases and multiple breaks in {\em Swift} X-ray light curves (\citealp{Nousek}). In particular, discrepancies between X-ray and optical light curves of GRB afterglows and the rich array of light curve morphology including extended plateaux, rebrightenings and flares challenge existing GRB models (see for a review \citealp{Piranreview}). Explanations for these features include complex jet structure, energy re-injection due to late-time central energy activity, variable microphysics and off-axis emission, with a combination of effects likely in action in most afterglows.
However, these model variations are in principle non-degenerate: linear and circular polarimetry have the ability to separate out the
various models for the behaviour of the (early) afterglows in a
completely independent way, as these models have different implications for the magnitude and orientation of the optical polarisation, as well as for their time-dependence (e.g. \citealp{Rossi}).

The ground-breaking discovery and interpretation of sudden achromatic steepening in light curves at $\sim1$ day post-burst in pre-{\em Swift} GRBs - so-called jet breaks - provided the first convincing evidence of highly collimated ejection (eg. \citealp{Rhoads97,Rhoads99,Sari2}) and allowed jet opening angles and collimation-corrected energies to be derived (e.g. \citealp{Frail}).
Jet breaks were expected to be ubiquitous in {\em Swift} light curves, but the added complexity and multiple breaks in these light curves have made jet breaks difficult to identify unambiguously (e.g. \citealp{Curran}). In contrast, the linear polarisation around the time of the jet break is predicted to have a unique signature (\citealp{Sari,Ghisellini,Rossi}): At early times a distant observer located slightly off-axis will see the afterglow as a ring shaped source, which has strong polarisation in radial direction (assuming the magnetic field is compressed in the plane normal to the motion, i.e. ordered in the plane of the shock). In the received light, integrated over the ring, the polarisation will largely cancel out. However, as the fireball decelerates the size of the ring increases and at some point the edge of the jet is reached, at which point the polarisation does not completely cancel out anymore and linear polarisation is observed. As the ring expands further, the opposite edge of the jet is reached as well, giving rise to a second peak in the linear polarisation curve.
Therefore polarimetry provides a useful tool to unambiguously identify jet breaks from other light curve breaks, though measured polarisation curves of GRB afterglows have 
in some cases shown puzzling features (see e.g. the cases of GRBs 030329 and 020405 discussed further in Section 4.1). In addition, the linear polarisation properties probe jet internal structure, as different polarisation properties are predicted for `uniform' or `structured' jet energy distributions (Rossi et al. 2004,  see also Section 4.2.2).

It has been shown that if some fraction of the magnetic field in the shock has large-scale order, the light may also be circularly polarised (Masumiya
\& Ioka 2003; Sagiv, Waxman \& Loeb 2004). For the forward shock (in the ambient medium), most models predict 0.01$-$1\% circular polarisation during the first day (e.g. \citealp{masu}), depending on the
strength and order of the magnetic fields and the wavelength of observation. Part of this range is within reach of large telescopes, as we will demonstrate in Section 3.2.

In the \sw\ era few polarimetric measurements have been performed. Arguably the most successful have been the early time linear polarisation measurements
of the afterglows of GRBs 060418 and 090102,
performed with the RINGO polarimeter on the rapidly responding robotic Liverpool Telescope (\citealp{Mundell,Steele}), the latter GRB showing a linear polarisation of 10\% just 190 seconds after burst.
\cite{Steele} interpret the light curve properties and high degree of polarisation in GRB\,090102 - which was detected when emission from the reverse shock dominated the received optical light - as evidence for large-scale ordered magnetic fields in the expanding fireball.
This detection of significant polarisation at early times provides strong motivation for following the polarisation properties of the fading afterglow for many hours or days after the GRB. However, late-time polarimetric observations of GRBs (see \citealp{Lazzatireview} for an overview) show that the combination of fading afterglow and low levels of polarisation (few percent) require large telescopes to obtain polarimetric observations over multiple days to high precision ($\lesssim$0.3\%), using multiple filters to distinguish dust-induced from intrinsic afterglow polarisation. 
In this paper we present the first such dataset for a \sw\ burst, and the most extensive polarimetric dataset of any burst since GRB\,030329 (\citealp{Greiner}). 

This paper is the first in a series on the afterglow of GRB\,091018. In this first paper we will describe our dataset in detail, and compare the observed polarisation light curves with those of other (pre-{\em Swift}) bursts and theoretical models.
In a forthcoming second paper (Wiersema et al. 2012 in prep; hereafter called Paper 2) we will discuss the dust properties in GRB sightlines from a combination of multi-wavelength polarimetry, spectroscopy, host imaging and broadband afterglow spectral energy distributions.
In a third paper (Paper 3) we will model the afterglow physics of GRB\,091018 in greater detail and fit the polarisation data to more detailed models, in particular considering energy injection.

This paper is organised as follows: in Section \ref{sec:obs} we describe the polarimetry, spectroscopy and broadband photometry observations, the data reduction and data calibration techniques; in Section 3 we discuss the detected features in the data; and in Section 4 we compare the data to models and previous studies of afterglow polarimetric light curves. 
Throughout this paper we adopt a cosmology with $H_0 = 71$ km\,s$^{-1}$\,Mpc$^{-1}$, $\Omega_m = 0.27$, $\Omega_\Lambda = 0.73$.

\section{Observations}\label{sec:obs}
GRB\,091018 triggered the Burst Alert Telescope (BAT) onboard the \sw\ satellite at 20:48:19 UT on October 18, 2009 (trigger 373172; \citealp{Stamatikos}). The prompt emission shows the burst to likely belong to the class of long bursts (\citealp{shorts}), with a duration $T_{90} = 4.4 \pm 0.6$ s (\citealp{Markwardt}; \citealp{Ukwatta} report a spectral lag
of $143 \pm 297$ msec). An X-ray and optical afterglow were found by the \sw\ X-ray telescope (XRT) and UV-optical telescope (UVOT). 
\cite{Chen} reported an afterglow redshift of 0.971 shortly after.
Based on the initial brightness of the UVOT afterglow we activated our VLT polarimetry programme (programme 084.D-0949, PI Wiersema).

\begin{figure}   
\includegraphics[width=8.5cm]{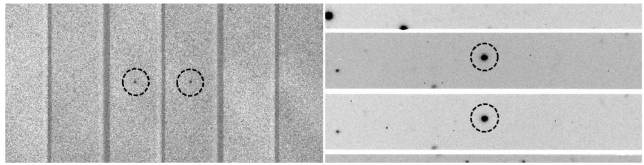}
\caption{Shown are small sections of dataframes from the ISAAC $Ks$ band polarimetry (Left; angle 0 frame from {\em linK1}, see Section 2.2) and optical FORS2 polarimetry (Right; angle -45 from circular polarimetry epoch {\em cir1}). The circles are 5 arcseconds in radius in each image, and show the source in the \ob\ and \eb\ beams.}
\label{fig_frames}
\end{figure}

\begin{table*}
 \centering
 \small
  \caption{Log of our polarimetric observations. The ID column gives a label to each polarimetric dataset to make discussion easier, with linear polarimetry datasets labelled ``lin'' and circular polarimetry ``cir''. The exposure time is the exposure time per retarder angle (4 angles for linear, 2 for circular polarimetry). The $Q/I$ and $U/I$ values are as measured: to obtain values corrected for scattering on Galactic dust, subtract the values derived in Section 2.1.3. \newline
  $^*$: combined value of three complete series (see Section 2.2). 
  }\label{table:polalog}
  \begin{tabular}{lllrrr} 
  \hline
  \multicolumn{6}{|c|}{FORS2, R$_{\rm special}$} \\
  \hline
ID         &    $T-T_0$     &  Exp. time      &$Q/I$                                         &  $U/I$                                   & $V/I$                        \\ 
             &   (mid; d)       &  (s)                   &                                                   &                                               &                      \\
 \hline
 {\em lin1}     &     0.13189   &     180.                              &    $-0.0014 \pm 0.0009$ & $-0.0036 \pm 0.0009$   &                      \\
 {\em cir1}     &     0.14302   &     300.                              &                                                  &                                                &  $-0.0005 \pm 0.0014$                 \\
 {\em cir2}     &     0.15133   &     300.                              &                                                  &                                                &   $0.0011 \pm 0.0015$                 \\
 {\em cir3}     &     0.15964   &     300.                              &                                                  &                                                &   $0.0003 \pm 0.0015$                  \\
 {\em cir4}     &     0.16795   &     300.                              &                                                  &                                                &  $-0.0017 \pm 0.0015$                  \\
 {\em lin2}     &     0.18043   &     240.                              &     $ 0.0009 \pm 0.0009$ & $-0.0040 \pm 0.0009$                 &                      \\                                    
 {\em lin3}     &     0.19701   &     300.                              &     $-0.0012 \pm 0.0008$ & $ 0.0024 \pm 0.0008$                 &                      \\                                    
 {\em lin4}     &     0.21282   &     300.                              &     $ 0.0010 \pm 0.0009$ & $-0.0023 \pm 0.0009$                   &                      \\                                    
 {\em lin5}     &     0.23030   &     300.                              &     $-0.0006 \pm 0.0009$ & $-0.0037 \pm 0.0009$                    &                      \\                                    
 {\em lin6}     &     0.24611   &     300.                              &     $ 0.0083 \pm 0.0009$ & $-0.0039 \pm 0.0009$                   &                      \\                                    
 {\em lin7}     &     0.26357   &     300.                              &     $ 0.0056 \pm 0.0009$ & $-0.0025 \pm 0.0009$                    &                      \\        
 {\em lin8}     &     0.27936   &     300.                              &     $ 0.0057 \pm 0.0009$ & $-0.0063 \pm 0.0009$                 &                      \\        
 {\em lin9}     &     0.45483   &     300.                              &     $ 0.0119 \pm 0.0012$ & $-0.0025 \pm 0.0012$              &                      \\        
 {\em lin10}   &    0.47064    &     300.                             &     $ 0.0067 \pm 0.0012$ & $-0.0009 \pm 0.0012$                &                      \\        
 {\em lin11}   &    1.13940    &     300.                              &     $-0.0162 \pm 0.0023$ & $ 0.0078 \pm 0.0023$               &                      \\        
 {\em lin12}   &    1.15522    &     300.                              &     $-0.0167 \pm 0.0022$ & $ 0.0260 \pm 0.0023$                &                      \\        
 {\em lin13}   &    1.17345    &     300.                              &     $ 0.0088 \pm 0.0023$ & $ 0.0130 \pm 0.0023$               &                      \\        
 {\em lin14}   &    1.18926    &     300.                              &     $-0.0123 \pm 0.0023$ & $-0.0147 \pm 0.0023$                &                      \\        
 {\em lin15}   &    1.20729    &     300.                              &     $-0.0070 \pm 0.0024$ & $-0.0054 \pm 0.0024$                &                      \\        
 {\em lin16}   &    1.22310    &     300.                              &     $-0.0126 \pm 0.0024$ & $-0.0080 \pm 0.0024$                &                      \\        
 {\em lin17}   &    1.36006    &     600.                              &     $ 0.0022 \pm 0.0018$ & $-0.0056 \pm 0.0018$              &                      \\        
 {\em lin18}   &    1.39183    &     600.                              &     $ 0.0014 \pm 0.0018$ & $ 0.0057 \pm 0.0019$             &                      \\        
 {\em lin19}   &    1.44926    &     600.                              &     $-0.0141 \pm 0.0019$ & $-0.0031 \pm 0.0019$            &                      \\        
 {\em lin20}   &    2.39023    &     720.                              &     $ 0.0111 \pm 0.0023$ & $-0.0092 \pm 0.0023$         &                      \\        
 \hline
   \multicolumn{6}{|c|}{ISAAC, Ks} \\
\hline
{ \em  linK}$^*$  & 0.4309    &  720.$^*$                        &   $0.0204 \pm 0.0078$         & $0.0008 \pm 0.0083$                  &                     \\
 \hline
\end{tabular}
\normalsize
\end{table*}

\subsection{$R$ band linear- and circular polarimetry}
\subsubsection{Data acquisition}\label{sec:linacq}
We acquired imaging polarimetry using the Focal Reducer and low dispersion Spectrograph (\ft) mounted on Unit Telescope 1 (Antu) of the VLT, using the \ft\ R$_{\rm special}$ filter.
After passing through a half- or quarter wavelength plate (in the case of linear and circular polarimetry, respectively), a Wollaston prism splits the incoming light into a ordinary and extraordinary beam (hereafter the \ob\ and \eb\ beam) that are perpendicularly polarised. These \ob\ and \eb\ beams are imaged simultaneously, using a mask to avoid overlap of the two beams on the chip, as shown in Figure \ref{fig_frames}. For each linear polarisation data set we used four rotation angles of the half wavelength plate, of 0, 22.5, 45 and 67.5 degrees.
The circular polarisation data were taken with -45 and +45 degree angles of the quarter wavelength plate.  

To ensure relatively homogenous polarimetric uncertainties as a function of time, exposure time was increased as the source faded. We employed a small dithering pattern, while taking care not to position the afterglow too close to the edges of the mask. We acquired a total of 20 linear polarisation series and 4 circular polarisation series, see Table \ref{table:polalog}.  

We began our \ft\ polarimetric monitoring with a single short sequence of linear polarimetry. After this, to make full use of the brightness of the afterglow we switched to circular polarimetry, for which
the models predict a much lower degree of polarisation, requiring a large number of detected photons. Directly following the circular polarisation, we continued our linear polarimetry. We obtained a further 6 sequences, and a further 2 at the end of the night. The next night we
obtained further datasets at the beginning and the end of the night. A last, deep, point was obtained in the third night. Note that for all \ft\ polarimetry we position the afterglow close to the centre of the 
\ft\ field of view, on chip 1, where instrumental polarisation is very low (well below our statistical errors, \citealp{patat}).

\subsubsection{Data reduction and calibration}\label{sec:linRreduc}
We bias and flat field corrected all data using tasks in IRAF and using standard twilight sky flats.  
We use our own software in combination with IRAF routines to perform aperture photometry on the \ob\ and \eb\ images of the afterglow to find the source fluxes $f_{o}$ and $f_{e}$, using the same approach as \cite{Rol}. We used a seeing matched aperture of
1.5 times the on-frame full width at half maximum (FWHM) of the point spread function, fitted per frame independently for the \ob\ and \eb\, as small differences in PSF shape may occur between \ob\ and \eb\ beams, particularly for objects far off-axis. The sky subtraction was done using an annulus of inner and outer radii  3 and 4 times the FWHM.
We measure $f_{o}$ and $f_{e}$ for all point-like objects in all frames. 

\begin{figure}   
\center
\includegraphics[width=6cm]{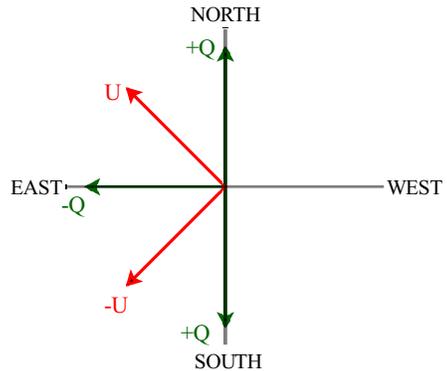}
\caption{Coordinate definitions used in this paper.}
\label{fig_coords}
\end{figure}

We express the polarimetry information in terms of the Stokes vector $\overrightarrow{S} = (Q, U, V, I)$ (see e.g. Chandrasekhar 1950), where the components of this vector have the following 
meaning: $Q$ and $U$ contain the behaviour of the linear polarisation; $V$ the circular polarisation and $I$ is the intensity. Generally we will use the normalised Stokes parameters
$q = Q/I$, $u = U/I$ and $v = V/I$ in this paper. Theoretical models are often expressed in terms of the polarisation degree $P$ and polarisation angle $\theta$. These quantities can be found from the Stokes parameters
through the relations:
\[
P_{\rm lin} = \frac{\sqrt{Q^2 + U^2}}{I}
\]
\[
\theta = \frac{1}{2}{\rm arctan}(U/Q) + \phi
\]
\[
P_{\rm cir} = V/I
\]
Constant $\phi$ is an offset defined so that the resulting angle $\theta$ conforms to the standard definitions of position angle (angle from North, counterclockwise, see Fig.\ref{fig_coords}):
\[ \phi = \left\{ \begin{array}{ll}
         0^{\circ} & \mbox{if $Q > 0$ and $U \geq 0$};\\
        180^{\circ}  & \mbox{if $Q > 0$ and $U<0$};\\
        90^{\circ}  & \mbox{if $Q<0$}.\end{array} \right. \]

Note that the conversion from Stokes parameters to $P$ brings with it complications, discussed further below, so wherever possible we will work in Stokes parameter space.

To measure the Stokes $Q/I$ and $U/I$ parameters a measurement at two rotation angles suffices, but this introduces substantial systematic uncertainties, caused by imperfect flatfielding, imperfect
background subtraction and imperfect behaviour of the half wavelength plate and Wollaston prism. Using four rotation angles with a constant stepsize (in this case of 22.5 degrees) means that several of these systematic effects cancel out (in particular the effects of background subtraction and flatfielding), leading to greatly improved polarimetric error terms. With these angles one can write (Patat \& Romaniello 2006):
\[
q = Q/I = \frac{2}{N}\sum\limits_{i=0}^{N-1} F_{i} \cos\left(i\pi/2\right)\,{\rm and}\,\, u = U/I = \frac{2}{N}\sum\limits_{i=0}^{N-1} F_{i} \sin\left(i\pi/2\right), \,\,\, 
\]
where $N$ is the number of half wavelength plate positions, and $F_i$ is the normalised flux difference of a source in the \ob\ and \eb\ beams at the $i$-th angle
\[
F{_i} =  \left(f_{o,i} - f_{e,i}\right) / \left(f_{o,i} + f_{e,i}\right) =  \left(f_{o,i} - f_{e,i}\right) / I .
\]
For the circular polarimetry, taken with two angles of the retarder, we can simply write
\[
v = V/I =  \frac{1}{2}\left(F_{45} - F_{-45}\right).
\]

After measurement of the fluxes of the sources in the images we compute their observed Stokes parameters and their uncertainties using the expressions above. 
The afterglow is always positioned on nearly the same position on CHIP1 (barring a small dither of $\sim16$ pixels in Y direction in some epochs), on the optical axis. As such we expect the instrumental linear polarisation to be negligible; see Patat \& Romaniello (2006) for a thorough error analysis of FORS1, whose polarimetry optics were moved to the FORS2 instrument in 2009. 
Additionally, circular polarimetry Stokes $V/I$ has no instrumental polarisation on axis.
In conclusion, the values of the Stokes parameters found through the methods above allow an investigation of the time dependence of the afterglow $q, u$ and $v$. 

Theoretical models are generally expressed in terms of the linear and circular polarisation fractions $P_{\rm lin}$ and $P_{\rm cir}$, and we therefore convert the Stokes parameter information
to these quantities using the equations above. We note that the uncertainty on the linear polarisation angle is a function of the intrinsic polarisation degree ($\sigma_\theta = \sigma_{P_{\rm lin}} / 2P_{\rm lin}$), which means that for the low polarisation values and faint fluxes of afterglows the uncertainties in $\theta$ are very large. Finally, the position angles found are corrected for the instrumental zero angle offset, which is $-1.48^{\circ}$ at 655 nm (the central wavelength of the filter)\footnote{Tabulated at the FORS instrument webpages}.

Errors on $q$ and $u$ are generally distributed as a Normal distribution and the Stokes parameters can have positive and negative values. In contrast, $P_{\rm lin}$
is a positive definite quantity. As demonstrated in \cite{Simmons} (their Section 2), integrating the equation for the distribution of measured ($P,\theta$) over $\theta$ shows that $P_{\rm lin}$ follows a Rice probability distribution rather than a Gaussian one. Directly using the equations above will lead to overestimated $P_{\rm lin}$ and incorrect confidence intervals, an effect generally referred to as the linear polarisation bias.  
The correction to the resulting $P_{\rm lin}$ and associated confidence ranges has been studied through both analytical and numerical (Monte Carlo) methods. Generally speaking, this correction
depends on $\sigma_P$, $P$, and the signal to noise (SNR) of $f_e + f_o$ (i.e. the SNR of $I_i$). In the literature one often sees the Wardle \& Kronberg (1974) prescription used, in which the input $P$ values are multiplied by
$\sqrt{1 - (\sigma_P / P)^2}$ to find the bias-corrected polarisation. We follow Sparks \& Axon (1999) in using a parameter $\eta = P \times {\rm SNR}_F$ to trace the expected behaviour of the bias and $\sigma_P$: when $\eta > 2$ the Wardle \& Kronberg correction is valid, and $\sigma_P$ is as computed directly from the uncertainty of $F$. In our FORS2 dataset before correction for Galactic polarisation (Section \ref{sec:linRGIP}) all data points have $\eta >2$, and the bias correction is in all cases small compared to $\sigma_P$.

\subsubsection{Polarisation induced by the Galactic interstellar medium}\label{sec:linRGIP}
To bring the afterglow $q$ and $u$ to an absolute frame, we need to correct for the linear polarisation induced by scattering on dust in our Galaxy (Galactic interstellar polarisation, GIP). 
Note that $V/I$ does not require a correction as there is no circular GIP and the resulting linear polarisation is much smaller than the threshold for noticeable linear - circular crosstalk (\citealp{patat}).

We assume here that the average intrinsic polarisation of a sufficiently large sample of field stars is zero, that the bulk of the polarising dust is between these stars and the observer, and that therefore the observed distribution of $q$ and $u$ of field sources measures the GIP.
We measure the Stokes $q$ and $u$ values of all sources that have a SNR above 20 in the FORS2 field, on both chips, and in all epochs. 
A large number of artefacts are visible in our FORS2 imaging polarimetry, which have the appearance of a group of stars, and which are likely caused by reflections at the retarder
plate\footnote{{\tt http://www.eso.org/observing/dfo/quality/FORS2/qc/\\problems/problems\_fors2.html}}; 
we take special care to avoid stars close to these artefacts.

We then eliminate sources that appear significantly extended (galaxies), which may give spurious linear polarimetric signals if seeing conditions change during a polarimetric sequence. We further eliminate sources that are close to the mask edges, and sources that are close to saturation in one
or more of the images.

It is well known that FORS1, whose polarimetric optics are now cannibalised in FORS2, displayed a roughly radial instrumental linear polarisation pattern, with
polarisation magnitude depending on the radial distance to the optical axis (see Patat \& Romaniello 2006). The shape of the linear polarisation pattern and its magnitude depend on the wavelength, and was found to reach nearly $\sim 1.5\%$ at the edges of the field of view (Patat \& Romaniello 2006).
While this effect is well calibrated for FORS1 in $V$ and $I$ band, a similar analysis has not yet been performed for FORS2. We therefore use a conservative approach and choose only stars close (radially) to the GRB position.
It is of some importance to not select only the brightest stars (which would give smallest uncertainty in the GIP value), as this may result in a bias towards the nearest stars (in distance), and therefore may not sample the complete Galactic dust column towards the GRB. We also make sure stars are selected with a fairly uniform distribution
over the field (in terms of position angle).

\begin{figure}   
\includegraphics[width=7cm]{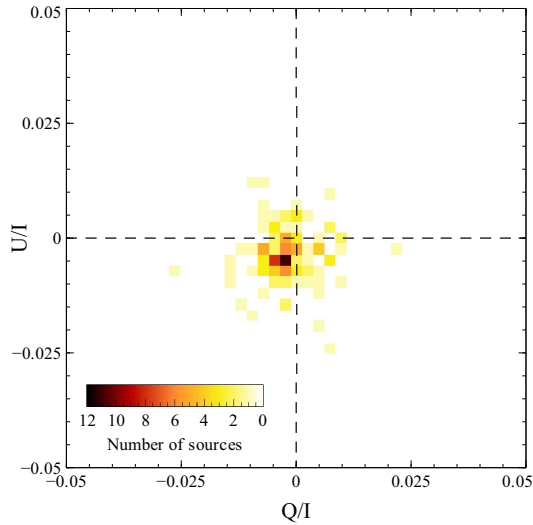}
\caption{Measurements of $q$ and $u$ of field stars to determine the contribution to the received polarisation by scattering on dust within our Galaxy. Dashed lines indicate $q=0$ and $u=0$. Binsize used in this plot is 0.0025.}
\label{fig_gip}
\end{figure}

We determine the GIP $q$ and $u$ values using three methods: we find the weighted mean of the field star distribution; we fit 1-dimensional Gaussian curves on the histograms of the $q$ and $u$ distributions; and we fit a 2-dimensional Gaussian function (with its orientation as a free parameter) on the $(q,u)$ distribution after converting this to a 2-dimensional histogram (where a range of bin sizes are used based on the number of input data points). 
We perform these fits for a range of limit values on the maximum radial distance $r$ between the input stars and the GRB; and on the maximum allowed $\sigma_P$ of the stars. 
Unfortunately there are only a small number of fairly bright stars near to the GRB position, making it necessary to use a large limit on the radial distance to obtain a sufficient number of datapoints.
We find a best balance between the number of sources in the fit and the uncertainties on the resulting $q_{\rm GIP}$ and $u_{\rm GIP}$ for limits $\sigma_P \lesssim0.35\%$ and $r \lesssim2.0$ arcmin (103 datapoints). Note that $r=2$ arcmin would correspond to $\sim0.23$ and $\sim0.34$\% instrumental polarisation in $V$ and $I$ band in FORS1, respectively; \cite{patat}.
Using these limits the three methods give very similar GIP values. 
In Figure \ref{fig_gip} the distribution of datapoints (103 points) is shown. We find $q_{\rm GIP} = -0.0028$ and $u_{\rm GIP} = -0.0036$, or $P_{\rm GIP} = 0.45$\%, $\theta_{\rm GIP}= 116^{\circ}$.
The 2-dimensional fit gives the width of the Gaussian along the $q$ axis as $\sigma_q = 0.0030$ and along the $u$ axis as $\sigma_u = 0.0025$, and the rotation of the containment ellipse from the $q$ axis in radians, counter-clockwise, as $\Theta = 1.7$ radians. The evidence for deviation from radial symmetry is not very strong,
but the analysis of \cite{patat} shows that the FORS1 off-axis polarisation pattern exhibited non-radial behaviour in the B band.
We use this ellipse as a conservative error in the GIP Stokes parameters, following \cite{Rol}.
The fairly large uncertainty in these is likely dominated by the contribution of the unknown instrumental polarisation pattern and the low number of bright stars within $\sim1$ arcminute of the GRB position.

The empirical relation  $P_{\rm GIP} \lesssim 0.09 E(B-V)$  (\citealp{Serkowskietal}) gives  $P_{\rm GIP} \sim0.3$\%, which compares well with the value derived above.

We now subtract from the GRB $q$ and $u$ values the $q_{\rm GIP}$ and $u_{\rm GIP}$ found above, to remove the (constant in time) component to the polarisation caused by the Galactic interstellar medium. We propagate the errors on the GIP Stokes parameters into the resulting corrected values $q_{\rm GIPcorr}$ and $u_{\rm GIPcorr}$. As is clear from Table \ref{table:polalog}, the afterglow polarisation at very early times (e.g. {\em lin1, lin3}) all but vanishes, indicating that at early times the detected polarisation is dominated by GIP. In many epochs, the $q_{\rm GIPcorr}$ and
$u_{\rm GIPcorr}$  errors are dominated by the uncertainties in the GIP Stokes parameters.
We compute GIP corrected values for the polarisation and position angle, which will bring the measured values onto an absolute scale. It is clear that the errors $\sigma_P$ are increased dramatically, while
$P$ decreases in several epochs (to near zero in e.g. {\em lin1}), i.e. a drop in $P/\sigma_P$. At the same time, the statistical detection signal to noise (i.e. SNR of $I_i$, simply referred to as SNR in Section \ref{sec:linRreduc}) stays the same. As the polarisation bias is a function of both SNR and $P/\sigma_P$, the polarisation bias correction is much larger in several epochs after GIP correction, showing
the necessity of treating GIP effects in $q,u$ space rather than $P,\theta$ space in low polarisation regimes. In several cases the Wardle \& Kronberg prescription is not valid (see \citealp{Simmons}), and we use the methods recommended by \cite{Simmons} for low $P/\sigma_P$ situations (which uses a maximum likelihood estimator, this estimator has lowest bias factor and lowest risk function values in this signal to noise range). In two cases this led to robust zero polarisation values. We refer to
\cite{Simmons} for a discussion on the probability of correctly inferring zero polarisation for different estimators. 
 
All resulting values are listed in Table \ref{table:polalogPPA}.

\begin{table}
 \centering
 \scriptsize
  \begin{tabular}{lllll} 
  \hline
ID                &   $P_{\rm lin}$    & $\theta$                    & $P_{\rm lin, GIPcorr}$ & $\theta_{\rm GIPcorr}$\\
                    &   (\%)                    &  (degrees)                & (\%)                                & (degrees)               \\
                    \hline
{\em lin1}   & $0.36 \pm 0.13$& $126.1 \pm  19.3 $ & $0\,\, (0.32)^\dagger$        &                                \\  
{\em lin2}   & $0.38 \pm 0.12$& $143.1 \pm  17.7 $ & $0.21 \pm 0.31$ & $177.0 \pm 47.5$\\  
{\em lin3}   & $0.24 \pm 0.11$& $60.2  \pm  25.1 $  & $0.56 \pm 0.27$ & $37.7 \pm 24.7$  \\  
{\em lin4}   & $0.22 \pm 0.12$& $148.5 \pm  28.4 $ & $0.26 \pm 0.31$ & $9.2 \pm 43.7$   \\   
{\em lin5}   & $0.35 \pm 0.12$& $132.1 \pm  19.0 $ & $0\,\, (0.32)^\dagger$        &\\  
{\em lin6}   & $0.90 \pm 0.12$& $168.9 \pm  7.9  $  & $1.07 \pm 0.30$ & $179.2 \pm 16.1$ \\  
{\em lin7}   & $0.59 \pm 0.12$& $169.5 \pm  12.0 $ & $0.78 \pm 0.31$ & $3.9 \pm 21.2$\\  
{\em lin8}   & $0.84 \pm 0.13$& $157.4 \pm  8.8  $  & $0.84 \pm 0.30$ & $171.1 \pm 19.9$\\  
{\em lin9}   & $1.20 \pm 0.17$& $175.5 \pm  8.1  $  & $1.44 \pm 0.32$ &$2.2 \pm 12.6$\\  
{\em lin10} & $0.65 \pm 0.17$& $177.7 \pm  14.7 $& $0.94 \pm 0.32$ & $8.0 \pm 18.6$\\  
{\em lin11} & $1.77 \pm 0.32$& $78.6  \pm  10.2 $  & $1.73 \pm 0.36$ & $69.8 \pm 11.7$\\  
{\em lin12} & $3.07 \pm 0.32$& $62.8  \pm  5.9  $   & $3.25 \pm 0.35$ & $57.6 \pm 6.1$\\  
{\em lin13} & $1.53 \pm 0.32$& $29.5  \pm  11.9 $  & $1.99 \pm 0.35$ & $27.6 \pm 10.0$\\  
{\em lin14} & $1.89 \pm 0.32$& $116.4 \pm  9.8  $  & $1.42 \pm 0.36$ & $114.6 \pm 14.0$\\  
{\em lin15} & $0.81 \pm 0.33$& $110.2 \pm  21.7 $ & $0.27 \pm 0.38$ & $101.6 \pm 47.1$\\  
{\em lin16} & $1.45 \pm 0.33$& $107.7 \pm  13.0 $ & $1.00 \pm 0.38$ & $102.2 \pm 20.2$\\  
{\em lin17} & $0.54 \pm 0.25$& $147.0 \pm  24.0 $ & $0.41 \pm 0.34$ & $168.8 \pm 36.6$ \\  
{\em lin18} & $0.52 \pm 0.26$& $39.5  \pm  25.3 $  & $0.97 \pm 0.32$ & $32.8 \pm 17.8$\\  
{\em lin19} & $1.42 \pm 0.27$& $97.6  \pm  10.7 $  & $1.08 \pm 0.35$ & $88.7 \pm 17.9$\\  
{\em lin20} & $1.40 \pm 0.32$& $161.6 \pm  13.1 $ & $1.45 \pm 0.37$ & $169.0 \pm 14.3$\\
\hline
{\em linK}$^*$ &  $2.0 \pm 0.8$& $10.9 \pm 20.9$                                   & $2.0 \pm 0.8$& $10.9 \pm 20.9$  \\
\hline
\end{tabular}
\normalsize
  \caption{Linear polarisation values, where coordinate definitions are used as in Figure \ref{fig_coords}. All values have been corrected for polarisation bias (after correction for GIP in the case of the last two columns), see Section  \ref{sec:linRGIP}. The uncertainties in the GIP $q,u$ value are fully propagated, and dominate the uncertainties in the GRB afterglow GIP-corrected $q,u$. 
  The GIP corrected polarisation is very low (with inferred values of zero in two cases), and as such the errors on $\theta$ are very large (as $\sigma_\theta$ is a function of $1/P$, see Section \ref{sec:linRreduc}).\newline
  $^*$: Combined value from {\em linK1, linK2} and {\em linK3}. The GIP correction in $K$ band is more than an order of magnitude smaller than the statistical error in the polarisation, so no correction is made.\newline
$^\dagger$: See Simmons \& Stewart (1985) for the probability of correctly inferring true polarisation $p_0 = 0$ from measured polarisations $P$ using a maximum likelihood estimator. The number in brackets is the 67\% confidence interval boundary.
} \label{table:polalogPPA}

\end{table}

\begin{figure}   
\includegraphics[width=8cm]{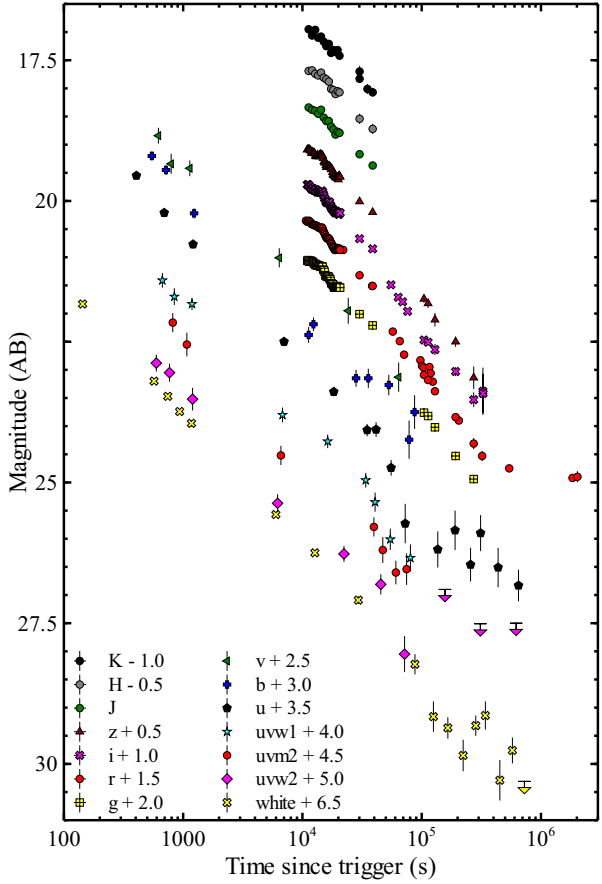}
\caption{Optical light curves used in this paper (for their values, see Table 4). All magnitudes are in the  AB photometric system, and the FORS2 R$_{\rm special}$ magnitudes are converted to $r$. Light curves in different bands are offset for clarity. }
\label{fig_optlc}
\end{figure}

\begin{figure}   
\includegraphics[width=8cm]{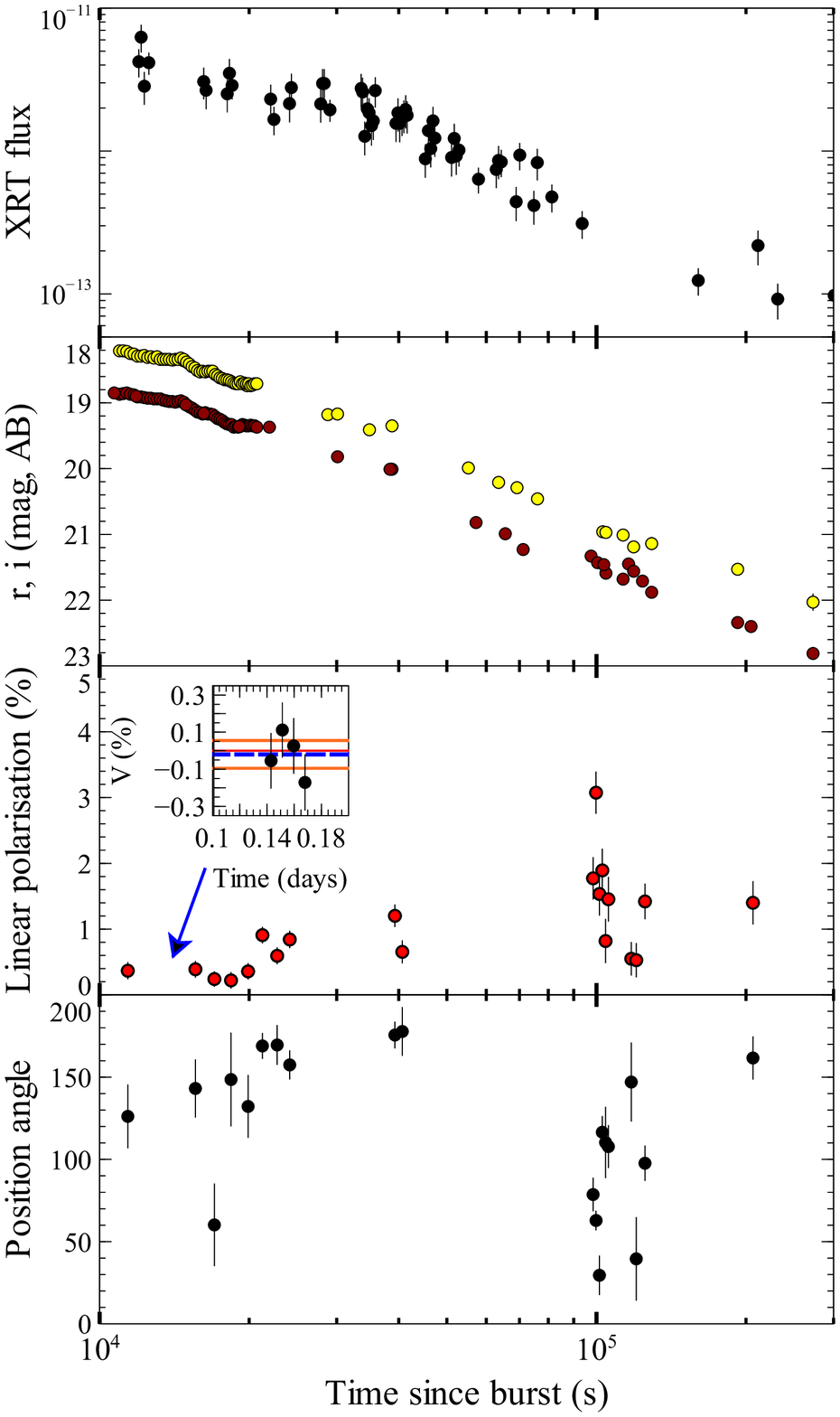}
\caption{
This figure shows the linear polarimetry light curve with broadband light curves as reference. \newline
The top panel shows the XRT X-ray light curve over this interval.
The panel below that shows the $r$ (red) and $i$ (yellow) light curves over this time interval ($i$ magnitudes shifted by -0.5 magnitudes). The lower two panels show the linear polarisation and polarisation position angle. We plot the polarimetric datapoints as observed, i.e. before correction for Galactic dust induced polarisation. The inset shows the circular polarimetry (four measurements of Stokes $V/I$) performed in between {\em lin1} and {\em lin2}. 
The horizontal thin red line shows the $V/I = 0$ level, and the blue dashed line shows the average of the four datapoints, $V/I = -0.0002$, with the solid orange lines indicating the error $0.00075$; leading to a 2$\sigma$ limit $ 0.15\%$. }
\label{fig_lc_pol}
\end{figure}

\begin{figure}   
\includegraphics[width=8cm]{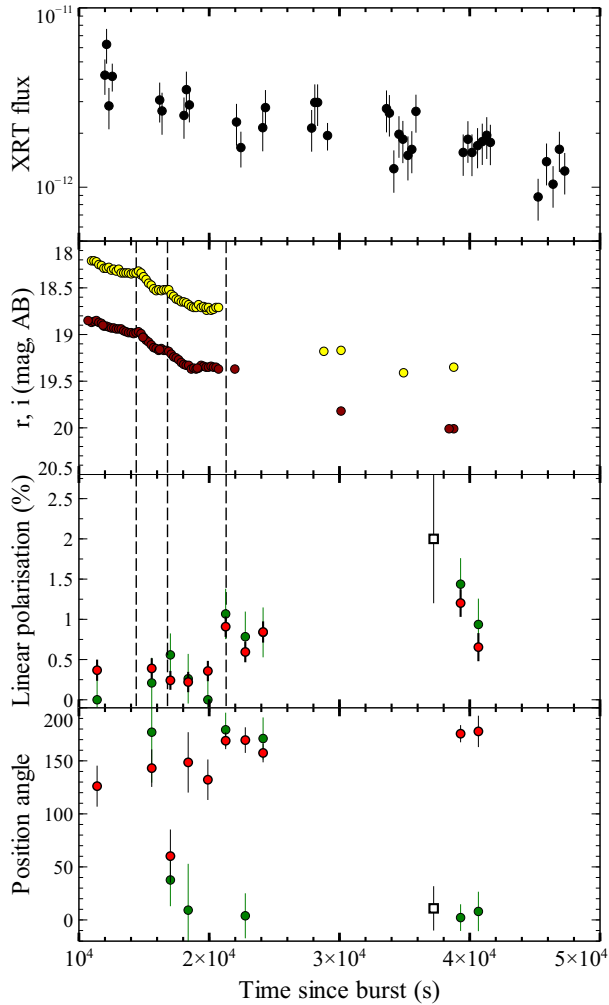}
\caption{Panels are as in Figure \ref{fig_lc_pol}, showing the behaviour of the source during the first night of polarimetric observations ({\em lin1 -- lin10}) - note the linear time axis. The lower two panels show the linear polarisation and polarisation position angle, but this time we plot both the polarimetric datapoints as observed (red) and after correction for Galactic dust induced polarisation (green). 
Open black squares give the ISAAC $Ks$ band polarisation degree and angle.
 Data points with zero GIP corrected polarisation have no associated position angle measurement. We note that position angles 
range [0--180], i.e. one can add 180 degrees to green datapoints with low angle values (to any datapoint one can add or subtract 180 degrees) to visually verify they behave similarly to the trend seen in the data before GIP correction.
Dashed vertical lines indicate the centres of three bumps found in the optical light curve (fits are in Section \ref{sec:lc}).
}
\label{fig_linnight1}
\end{figure}

\begin{figure}   
\includegraphics[width=8cm]{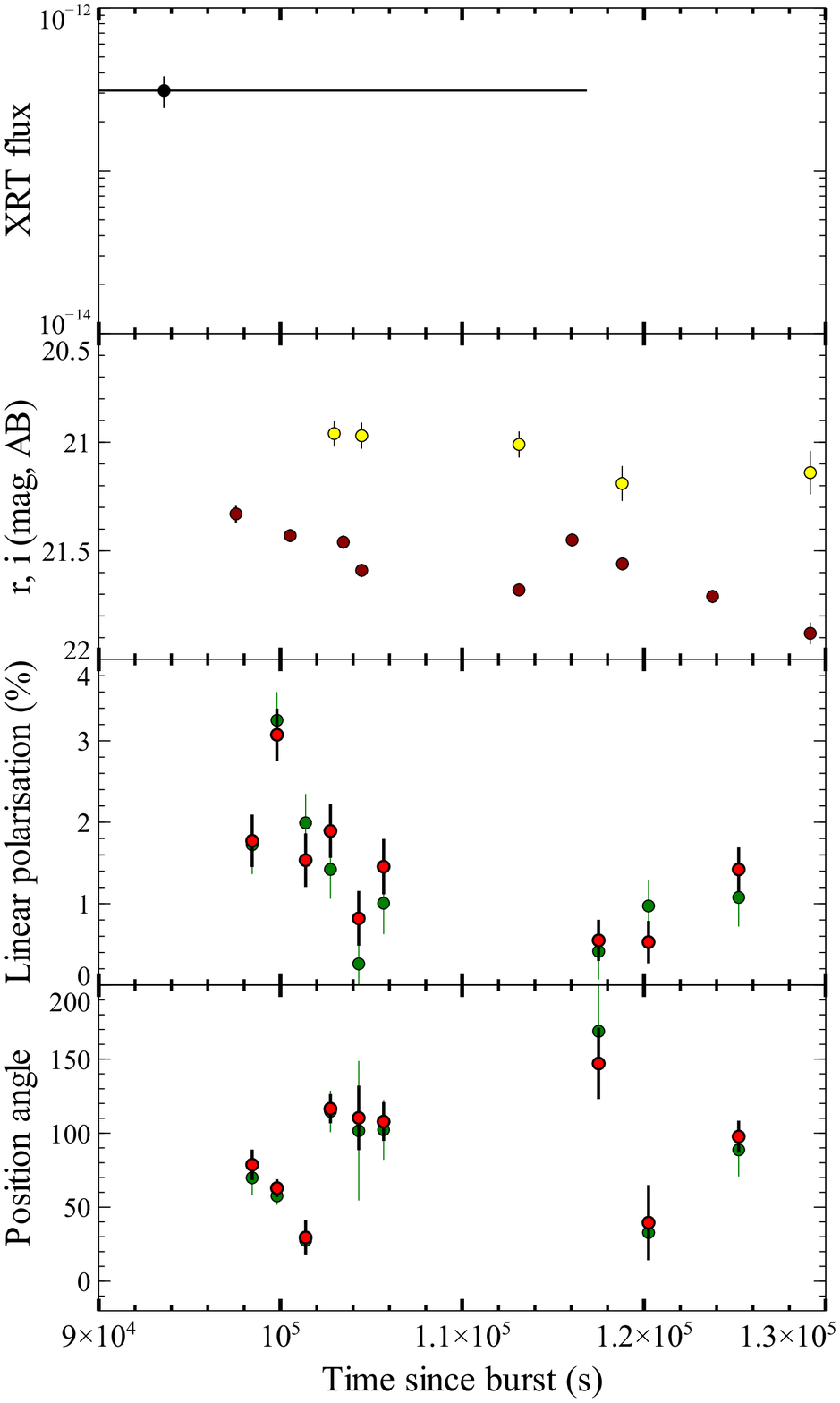}
\caption{Panels are as in Figure \ref{fig_lc_pol}, this time showing the behaviour of the source during the second night of polarimetric observations ({\em lin11 -- lin19}). A 180 degree rotation of the position angle is apparent, sampled in the first 6 datapoints. }
\label{fig_linnight2}
\end{figure}

\subsection{K band linear polarimetry}
\subsubsection{Data acquisition}
We obtained an imaging polarimetry data set using the short wave polarisation (SWP) mode of the ISAAC instrument mounted 
on Unit Telescope 3 (Melipal) of the VLT, using the $Ks$ filter. Observations coincided in part with $R$ band linear polarimetry, see Table \ref{table:polalog}.
ISAAC SWP mode uses a similar approach to FORS2 to obtain polarimetry: a Wollaston prism is inserted in the light path,
and the incoming light is split into a ordinary and extraordinary beam (the \ob\ and \eb\ beam) that are perpendicularly polarised.
The \ob\ and \eb\ beams are imaged simultaneously, using a mask to avoid overlap. The angular separation of the \ob\ and \eb\ images is wavelength dependent.
The main difference between ISAAC and FORS2 imaging polarimetry is that a half wavelength plate is not present in ISAAC, requiring rotation of the instrument to
obtain observations at the necessary angles; and the location of ISAAC in the Nasmyth focus. 

At each instrument angle we use integration times of $4 \times 15$ seconds, executed 4 times with significant dithering to facilitate
sky subtraction. 
Because of the aperture mask, less than half of the field of view is visible at any given rotation angle, and one generally uses a dither pattern designed to sample the entire field. However,
as the afterglow is faint and polarisation low, we dither such that the afterglow always stays within the mask opening.  
After these $4 \times 60$ seconds exposure, the instrument is rotated, and observations are repeated for the new position angle. 
By mistake we employed cumulative rotation offsets of 0, 22.5, 22.5 and 22.5 degrees. As ISAAC does not have a halfwavelength plate, the better choice of 
cumulative angles would have been 0, 45, 45, 45, so as to achieve the same effects as in the FORS data (switching of the \ob\ and \eb\ images to minimize flatfield errors
and Wollaston throughput uncertainties). These series of observations are repeated three times, resulting in complete sets {\em linK1, linK2} and {\em linK3}. 
Note that because of the rotation of the instrument, field stars are only visible in some of the rotation angles, and rotate behind
the mask at other angles. We note that a series of data were taken before {\em linK1} with an erroneous dithering pattern: the afterglow disappeared behind the
aperture mask at the third and fourth angle. In further analysis we will only use the complete cycles {\em linK1 - linK3}: the calibration as described below requires
measurement of both $q$ and $u$ for a given parallactic angle.

\subsubsection{Data reduction and calibration}
Data were corrected for dark current, flat fielded and background subtracted using tasks in IRAF, and the 4 exposures of 60 seconds per angle per cycle were co-added. 
Aperture photometry was performed on the \eb\ and \ob\ images of the afterglow, in the same way as described above for FORS2.  
We compute the normalised Stokes parameters:
\[
q = Q/I = \frac{(I_{o,0} - I_{e,0})}{(I_{o,0} + I_{e,0})} \,\,\,\,\,{\rm and}\,\,\,\,\, u = U/I = \frac{(I_{o,45} - I_{e,45})}{(I_{o,45} + I_{e,45})},
\]
using the data with rotation angle 0 and 45 degrees. As we did not take data with 90 and 135 degree angles, we can not use these to cancel out systematics as we did with the FORS2 data.
We expect these effects to be of the same order or smaller as the statistical errors (the afterglow was faint in $K$). 
We also compute Stokes parameters from the 22.5 and 67.5 degree data:
\[
q_2 = (Q/I)_2 = \frac{(I_{o,22.5} - I_{e,22.5})}{(I_{o,22.5} + I_{e,22.5})} \,\,\,{\rm and}\,\,\, u_2 = (U/I)_2 = \frac{(I_{o,67.5} - I_{e,67.5})}{(I_{o,67.5} + I_{e,67.5})},
\]
which we will calibrate independently (they differ from $(q,u)$ by a rotation in coordinate basis).
 
ISAAC is a Nasmyth focus instrument, and as such less well suited to accurate polarimetry: the instrumental polarisation of Nasmyth instruments is dominated by
reflection of the tertiary mirror M3 (at 45 degrees, therefore introducing strong polarisation which is highly dependent on position or parallactic angle), which requires particular care in calibration.
Our calibration strategy closely follows the recommendations of Joos et al. (2008) and Witzel et al. (2011).
These authors highlight in their papers the great benefits of using a train of Mueller matrices to create a polarimetric model of the instrument plus telescope for Nasmyth mounted instruments: all elements contributing to instrumental polarisation effects are described in $4\times4$ matrices acting on the Stokes vector $\overrightarrow{S}$. These matrices are then multiplied to give a Mueller matrix of the whole telescope/instrument for a given wavelength and a given set-up. We refer to Joos et al. (2008) for a detailed breakdown of the required matrix components. The Mueller matrix  ${\bf M}(\theta,\lambda)$ describing the effects of the M3 mirror on the polarisation is a function of the light incident angle $\theta$, wavelength $\lambda$ and the refraction index components $n$ and $k$ of the reflecting surface of M3. For these latter values we use the values at the $K$ band as derived by Witzel et al. (2011). 
 
We obtained observations of the unpolarised DA white dwarf GJ 915 (WD 2359-434; \citealp{fossati}) using the same instrument rotation angles as the science data and at a similar parallactic angle (60 degrees at mid of the observation vs. 50 degrees for the mid of the range of the science series). Our calibrator observations were obtained just before the ISAAC science observations.
We add to these observations from the ESO data archive of unpolarised white dwarf WD 1620-391 (using identical rotator angles as our data), taken on 16 September 2009, with parallactic angle 76 degrees. Using these stars and the Mueller matrices described above we calibrate our Stokes parameters. Using our own standard star observations and those of 2008 gave consistent results within the errors. Because of the way we dithered our data and the low number of bright field stars, we can not refine our calibration using field stars.
We then rotate the $(q_2, u_2)$ values to the same orientation as $(q,u)$. Because the source was fairly faint in $Ks$ band, the errors on the Stokes parameters of {\em linK1-3} are large, and we therefore
combine them together, finding a combined $q = 0.0204 \pm 0.0078$  and $u = 0.00081 \pm 0.0083$. 

There are no field stars available for an independent GIP calibration in $Ks$ band, but we can estimate the likely contribution of GIP to the observed values by using the approximate relation  
$p(\lambda) \propto \lambda^{-1.8}$ (see \citealp{Martin2}). While the value of the exponent has fairly large uncertainty, it is clear that a GIP polarisation in $R$ band of $\sim0.45$\% corresponds to 
an expected GIP contribution in the $Ks$ band of $<0.1$\%, i.e. an order of magnitude smaller than the statistical uncertainties in the value above. We therefore make no correction to the $Ks$ band Stokes parameters above. Using the same techniques as for the FORS2 data we compute a bias corrected linear polarisation in $Ks$ band of $2.0 \pm 0.8$\% with a position angle $10.9 \pm 20.9$ degrees.

\begin{figure}   
\includegraphics[width=7cm]{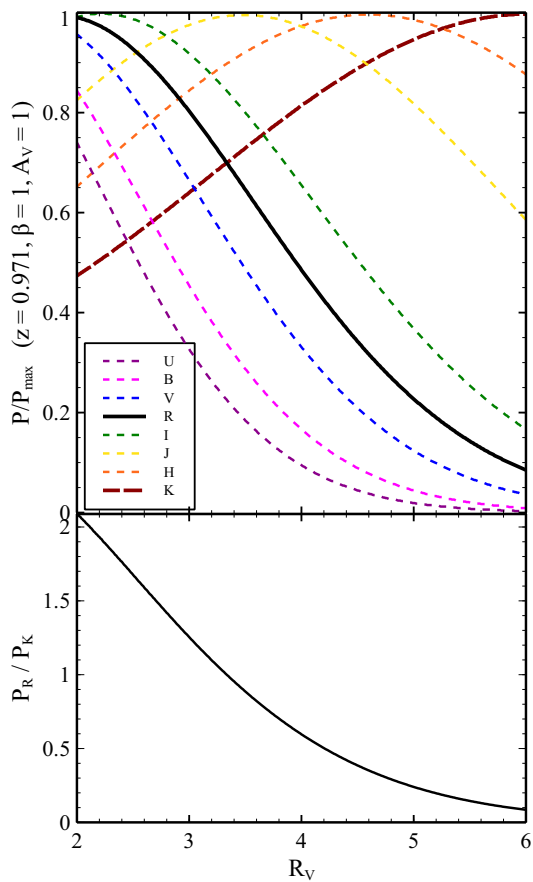}
\caption{Upper panel: $P/P_{\rm max}$ for a range of $R_V$ values, at the redshift of GRB\,091018 and using host galaxy extinction $A_V = 1$ magnitude. Lower panel: resulting ratio of $R$ band and $K$ band linear polarisation (in the case that all detected polarisation is due to dust scattering in the host galaxy). }
\label{fig_ppmax}
\end{figure}

\subsection{X-shooter spectroscopy}
After the discovery of an optical afterglow, we activated our VLT \xs\ (mounted on UT2 of the VLT, Kueyen)  programme for GRB afterglow spectroscopy (programme ID 084.A-0260, PI Fynbo). We obtained
$4 \times 600$s exposures with midpoint 0.1485 days ($1.283 \times 10^4$ seconds) after burst. 
\xs\ is an echelle spectrograph with three arms, the UVB, VIS and NIR arms, separated by dichroics, resulting in a wavelength coverage $\sim0.3 - 2.4 \mu$m. We used
a $1\times2$ binning in the UVB and VIS arms (binning in the wavelength coordinate). A 5 arcsecond nod throw was used to facilitate accurate sky subtraction,
particularly important in the NIR arm, resulting in exposures taken in a ABBA pattern.
We processed the data using version 1.3.0 of the ESO \xs\ data reduction pipeline (Goldoni et al. 2006), using the so-called physical mode. 
 
Flux calibration is achieved using exposures of the flux standard star BD+17 4708. Telluric line correction of the NIR arm spectrum was performed using methods outlined in Wiersema (2011) with the SpeXtool software (\citealp{SpeX}), using observations of the B9V star HD 16226. 
The resulting spectra are normalised.

\subsection{Optical and near-infrared photometry}
\subsubsection{GROND afterglow photometry}
The seven channel Gamma-Ray Burst Optical and Near-Infrared Detector (GROND; Greiner et al. 2008) mounted on the ESO 2.2m telescope observed the field of GRB\,091018 using $g', r', i', z', J, H, Ks$  filters beginning as soon as the source became visible from La Silla observatory (\citealp{GRONDGCN}). Photometric calibration was achieved through observations of photometric standard star fields. A spectral energy distribution using some of the GROND data presented here was published earlier in \cite{Greinerdark}.
 
\begin{figure}   
\includegraphics[width=8cm]{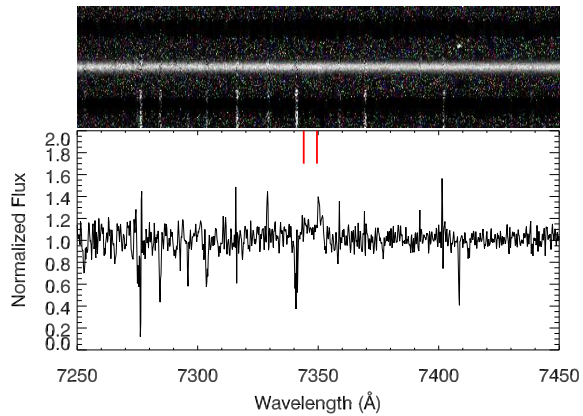}
\caption{Emission lines of the [O\,{\sc II}] doublet in the spectrum. The 2D spectrum is shown in the top panel. Weak residuals from imperfect subtraction of sky emission lines can be seen
as vertical stripes, also showing up as narrow residuals in the 1D spectrum.}
\label{fig:o2}
\end{figure}

\subsubsection{Faulkes Telescope South afterglow photometry}
We observed the optical afterglow using the Faulkes Telescope South (FTS), at Siding Spring, Australia.            
Observations were performed at two different epochs (about 15.3 hr and 3.7 d after the burst event), using $R$      
and {\it i} filters. These observations are complementary to those of GROND, filling the gap in coverage during         
daytime in Chile. All the images were cross-calibrated using the GROND field star measurement.

\subsubsection{VLT afterglow photometry}
As part of our linear and circular polarisation monitoring campaign described above, a large series of acquisition images were taken for accurate positioning
of the afterglow within the aperture mask. Within each of these images the afterglow is detected at good signal to noise levels.
We reduce the 14 FORS2 and 2 ISAAC acquisition images using flat fields, dark- and bias fields taken the same night. The FORS2 acquisition data were taken using the R$_{\rm special}$ filter, the
ISAAC data using the $Ks$ filter. We use a sample of field stars in the GROND data to bring the FORS2 acquisition photometry to the same homogeneous calibration. The ISAAC data were calibrated using four 2MASS stars in the field of view. The resulting magnitudes were converted to AB magnitudes, the resulting values are reported in Table \ref{table:photlog}.  

In addition to the polarimetry acquisition images, we also photometer the acquisition image taken for the \xs\ spectroscopy, calibrating onto the GROND system.

\subsubsection{Gemini-South afterglow and host photometry}
We followed the late-time afterglow with GMOS-South, mounted on the Gemini South telescope, Chile. We observed the source at three epochs, in Sloan r band (GMOS filter $r\_{\rm G0326}$), see Table \ref{table:photlog}.
We reduce the GMOS images using twilight flat fields and bias fields taken the same night using tasks from the {\sc Gemini} packages in {\sc IRAF}.
In all three data sets the host galaxy of the burst is visible. Photometric calibration is performed relative to the GROND calibration.

\subsubsection{UVOT afterglow photometry}
{\em Swift}'s UltraViolet-Optical Telescope (UVOT, \citealp{Roming}) began settled observations 68 seconds after the
BAT trigger, beginning with a 150 s exposure in the white filter followed by a 245 s exposure in the u band and
thereafter cycling through all 7 lenticular filters.       
We performed photometry and created light curves using the tool {\it uvotproduct} provided in the {\em Swift} software. The     
latest calibration data as of July 2011 were used. We binned the data requiring a minimum detection significance    
of 3$\sigma$ per bin; errors are given at the 1$\sigma$ level.

\section{Results}
\subsection{Light Curves and spectral energy distributions}\label{sec:lc}
\subsubsection{X-ray light curve fits}
The {\em Swift} XRT X-ray data, as analysed through the {\em Swift} XRT catalogue\footnote{\tt http://www.swift.ac.uk/xrt\_live\_cat/} (see Evans et al.~2009 for details), show an initial shallow decay phase, with
power law decay index $\alpha_1 = 0.41^{+0.06}_{-0.08}$. A break in the light curve at $587^{+85}_{-91}$ seconds sees the decay continue with a steeper index $\alpha_2 = 1.24\pm0.04$. 
Between $\sim10^4$ and $10^5$ seconds a broad bump is visible in the X-ray light curve. The {\em Swift} XRT catalogue lists the best fit solution ($\chi^2/{\rm dof} = 117/133$) using two breaks in quick succession, at $2.24^{0.39}_{-0.62} \times10^4$ seconds and $2.69^{+1.29}_{-0.16} \times 10^4$ seconds, with the first leading to a negative index (i.e. a brightening, but with poorly constrained index), and the last break leading to a final decay index of $\alpha_4 = 1.59^{+0.12}_{-0.11}$.
It is likely that the behaviour between $\sim10^4$ and $10^5$ seconds is the result of a system of late time flares (or some other form of short time scale rebrightening), in which case a more realistic alternative description may be that of a single break at $4.7^{+2.6}_{-1.0}\times 10^4$ seconds, with $\alpha_2=1.17 \pm0.03$ and $\alpha_3 = 1.54^{+0.18}_{-0.13}$. This gives a poorer  $\chi^2$ statistic ($\chi^2/{\rm dof} = 137/135$), as a cluster of datapoints interpreted as due to flaring/bump in this scenario are off the best fit. The X-ray light curve and the fit are shown in Figure \ref{fig:optical_fit}.

\subsubsection{Optical light curve fits}
\begin{figure}   
\includegraphics[angle=-90,width=8.5cm]{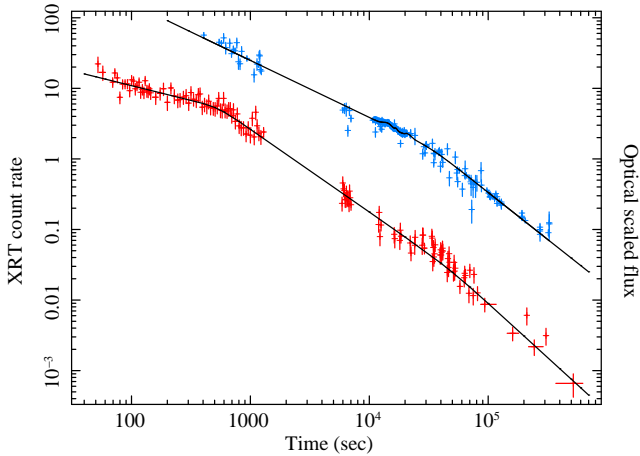}
\caption{The optical light curve, in arbitrary flux units as described in Section \ref{sec:lc}, is shown in blue (top), with the best fitting light curve superposed: a broken power law and three Gaussian components describing the behaviour of flares. The red datapoints show the X-ray light curve, in units of count rate, fit with a double broken power law, see Section \ref{sec:lc} for the fit parameters.}
\label{fig:optical_fit}
\end{figure}

The optical light curves presented in this paper have their densest coverage through the GROND observations, as described above. Figures \ref{fig_optlc}, \ref{fig_linnight1} and \ref{fig_linnight2} show the presence of bumps in the light curve, on top of the usual power law decay. Bumps in afterglow light curves are fairly common, though they require good signal to noise to be identified. Three distinct bumps, each lasting roughly half an hour, are readily visible in the first night data, and are indicated by dashed vertical lines in Figure \ref{fig_linnight1} (see Section \ref{sec:lc}; these bumps are also detected by the PROMPT telescopes, \citealp{PROMPT}). Similar to the X-rays, the presence of bumps makes the light curve fits somewhat complex: $\chi^2$ fit statistics are largely driven by datapoints belonging to bumps covered by GROND. To find a best possible way to discern bumps and power law breaks, we use the method described in \cite{Curran2}: data from all filters were combined through a free offset fit (i.e. we assume data from all filters have the same temporal decay but we make no assumptions regarding the relative offsets). Simulated annealing is used to fit the offsets, temporal indices and break times, and a Monte Carlo analysis with synthetic data sets is used for error determination: the normalisation error is added in quadrature to the datapoints in the resulting combined light curve.
We exclude very early data (i.e. the first UVOT datapoint at around 100 seconds), late time UVOT data ($>10^5$ seconds) and the late time Gemini data, which are nearly entirely host-dominated. The resulting shifted datapoints are placed on an arbitrarily scaled flux scale, and we fit this resulting dense light curve. We fit a broken power law, which gives a reduced $\chi^2 = 1.68$ for 365 degrees of freedom. We then add Gaussian functions one by one to empirically take care of the bumps in the light curve (i.e. this is not a physically motivated description of light curve bumps). The best fit is obtained using three Gaussians and a broken power law, and is displayed in Figure \ref{fig:optical_fit}, with 
reduced $\chi^2 = 1.12$ (356 degrees of freedom). 
We find best fitting parameters for the broken power law component of $\alpha_1 = 0.81 \pm 0.01$, $\alpha_2 = 1.33 \pm 0.02$ and a break time of $t_{\rm break} = 3.23 \pm 0.16 \times 10^4$ seconds. The three Gaussians have best fitting centre times $1.43 \pm 0.01 \times 10^4$, $1.68 \pm 0.01 \times 10^4$ and $2.13 \pm 0.06 \times 10^4$ seconds. Their widths as given by the standard deviation of the fitted Gaussians are $\sigma = 1140 \pm 200, 370 \pm 80$ and $1026 \pm 230$ seconds, respectively, though we note that the third bump is only partially covered by datapoints, resulting in the larger errors in central time and width. Adding the late time Gemini data to the dataset after subtracting the host brightness (the last Gemini epoch) does not change the fits.

\subsubsection{Spectral energy distributions and light curve interpretation}
\begin{figure}   
\includegraphics[width=8.5cm]{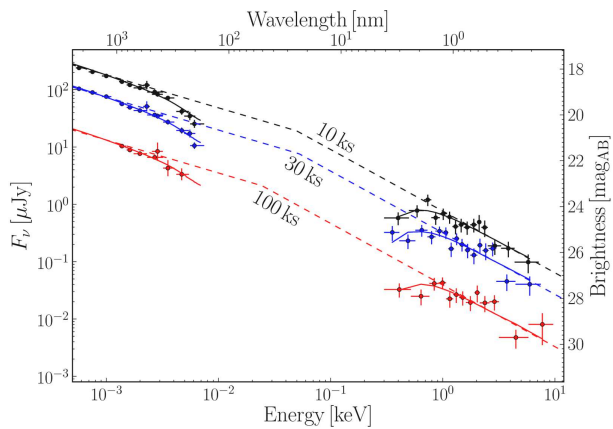}
\caption{X-ray - optical spectral energy distributions at three epochs (observer frame). Fits are done with a broken power law with a fixed difference between the two power law spectral indices, and with the same X-ray absorption and optical extinction for all three epochs. }
\label{fig:seds}
\end{figure}

In a study of the host galaxy dust extinction properties of sightlines towards GRB afterglows using single epoch broadband spectral energy distributions (SEDs), Schady et al. (2011) report on a late time SED of the afterglow of GRB\,091018. This SED (at $3\times10^4$ seconds, using GROND, UVOT and XRT data), is best fit with a broken power law, indicating the presence of a synchrotron cooling break in between X-ray and optical/UV wavelengths.  We extend this analysis, fitting the X-ray - optical SEDs at three representative epochs: at $10^4, 3\times10^4$ and $10^5$ seconds, using a fitting method described in detail in \cite{KruehlerSED}. 
We assume there is no change in X-ray absorption and optical extinction between the three epochs. We fit using a broken power law (with the two power law indices linked with $\beta_1 = \beta_2 - 0.5$) and SMC-like extinction law, shown by \cite{Schady} to be the best fit for this afterglow, with the $E(B-V)$ and $N({\rm H})$ fit simultaneously for all three epochs.  
Bands with a possible contribution from the Lyman limit are excluded from the fit. The fit statistics are computed using C statistics in the X-rays (see Evans et al.~2009) and
$\chi^2$ in the optical. The fits are displayed in Figure \ref{fig:seds}. The three SEDs are best fit ($\chi^2 = 27.04$ using 31 bins, C-stat $= 315.35$ using 433 bins) with the following properties (errors at 90\% level confidence): 
$E(B-V) = 0.024 ^{+0.005}_{-0.006}$, $N({\rm H}) = 1.7 \pm 0.8 \times 10^{21}$ cm$^{-2}$ and power law slope $\beta_1 = 0.58 \pm 0.07$. These values are in agreement with the results of \cite{Schady}. We find power law break
energies $E_{\rm break} = 0.05^{+0.1}_{-0.03}$, $0.05^{+0.1}_{-0.03}$ and $0.025^{+0.05}_{-0.017}$ keV for the SEDs at 10, 30 and 100 ksec, respectively.

We now compare the decay indices and spectral indices to closure relations for some simple models. The spectral indices found from the SED fits imply a
power law index for the electron energy distribution $p = 2.16 \pm 0.14$. If the blastwave propagates in a homogeneous density medium, the predicted afterglow decay index
for the regime $\nu < \nu_c$ is  $3(p-1)/4 = 0.87 \pm 0.11$, and for $\nu>\nu_c$ it is $1.12 \pm 0.11$. A wind-like medium (where density decreases with $r^2$) for $\nu < \nu_c$ would imply $\alpha = (3p-2)/4 = 1.37 \pm 0.11$; and for $\nu > \nu_c$ has the same index as for the homogeneous medium. 
The values for a homogenous medium agree well with the values determined from the light curve fits (using the two break model on the XRT data) for the 
pre-break slopes, and are consistent with the presence of a cooling break between X-ray and optical frequencies. In addition, in a wind-like medium the cooling break frequency should increase with time, which is not observed. 

The break time in optical and X-rays are broadly in agreement, and the SEDs (Fig \ref{fig:seds}; the first SED is before the break, the last after) show no evidence that the break is wavelength dependent (chromatic). As such, identifying this break with a jet break is compelling, particularly as this break is the second light curve break seen in the X-ray afterglow, with the first likely marking the end of energy injection. However, we note that the post-break decay indices ($1.33\pm 0.02$ and $1.54\pm 0.15$ for optical and X-rays, respectively) are too shallow compared to predictions from the standard fireball model, both in the case of
non-sideways spreading jets (which predicts post-break decay indices of $1.52 \pm 0.11$ for optical and $2.12 \pm 0.11$ for X-rays) and sideways spreading jets (which should give post-break decay indices of $p$).                                                                             

An alternative interpretation could be that the break is not a jet break, but rather a break marking the start of the ``normal'' afterglow phase, i.e. the decay indices after this break are pre-jet break indices, and the phase before this break is dominated by energy injection. In this case the post-break optical index of $1.33 \pm 0.02$ is consistent with the closure relations in case of a wind medium, which predicts $1.37 \pm 0.11$. However, in that interpretation the optical decay index should be steeper than the X-ray one, and the predicted X-ray decay index of $1.12 \pm 0.11$ is not consistent with the measurement.

We note that the homogeneous analyses of large samples of {\em Swift} XRT afterglows have shown that a large fraction of afterglows with candidate jet breaks (i.e. a light curve break significantly after a plateau phase) produce post-break decay indices that are too shallow compared to simple model predictions (\citealp{Willingale}; \citealp{Evans09}, their figure 10; and
\citealp{Racusin}, their figure 2). Both these papers further point out several additions to the models that can possibly explain some of this discrepancy, for example the presence of low-level continuous energy injection past the plateau phase, or peculiar jet structure and sightline configurations. In the following we will
consider the observed break a candidate jet break, and will discuss alternative options as well.

If the break is a jet break, we can correct the isotropic energy $E_{\rm iso}$ for beaming. Using  $E_{\rm iso} = 3.7 \times 10^{51}$ erg (\citealp{Konus}) and
making the same assumptions as in \citealp{Kocevski}, we find a jet opening angle $\theta_{\rm jet} \sim 0.059$ radians, or 3.4 degrees, leading to  
$\log(E_\gamma) \sim 48.9$. This is a fairly low value for $E_\gamma$, but not unprecedented: similar values are found for several other {\em Swift}
bursts (\citealp{Kocevski}) and pre-{\em Swift} bursts (\citealp{Frail}).

\subsection{Circular polarisation}\label{sec:circpol}
The presence of a (weak) ordered magnetic field in addition to a chaotic, incoherent magnetic field generated by the post-shock turbulence, has been proposed
(e.g. \citealp{Granot}) to explain the non-detection of the swing in linear polarisation position angle at the time of the jet break, expected in some models (e.g. Lazzati et al. 2003). In this picture, the observed polarisation may be dominated by the weak large scale ordered field, while emissivity is dominated by random tangled field made by the post-shock turbulence. Polarisation variability occurs as a result of changes in the ratio of the ordered-to-random mean-squared field amplitudes (Granot \& Koenigl 2003). This results in much weaker changes in the polarisation and polarisation position angle light curve around the jet break time. A direct test of this proposition can come from deep circular polarisation. Some searches have been performed at radio wavelengths during radio flares (Granot \& Taylor 2005), but with fairly poor sensitivity (the best upper limit on circular polarisation is 9\% for GRB\,991216). 

Our dataset contains four measurements of circular polarisation in optical, $R$ band, wavelengths, see Table \ref{table:polalog}, each with uncertainties of 0.15\%. Under the assumption that over the
interval that the data were obtained ($\sim0.7$ hours, see Table \ref{table:polalog}) there is no change in circular polarisation, we can combine these together. We find a combined value of the circular polarisation Stokes parameter of $v = V/I = -0.00020 \pm 0.00075$, which leads to a formal 
$2\sigma$ upper limit of $P_{\rm circ} < 0.15$\%: the deepest limit on circular polarisation of a GRB afterglow to date. 
Figures \ref{fig_optlc} and \ref{fig_linnight1} show that during the circular polarimetry epoch, the optical light curve shows a low amplitude bump. As such, the limit on the circular polarisation can be seen as a limit on the circular polarisation of the light of the bump plus that of the underlying afterglow.

\subsection{Interstellar polarisation within the host galaxy}
The variability of the detected linear polarisation indicates that the majority of detected polarisation can be associated with the afterglow. However, scattering of afterglow light onto dust grains within the host galaxy (HGIP (host galaxy interstellar polarisation), following terminology of \citealp{GorosabelSN}) can induce noticeable linear polarisation, depending on the dust scattering geometry and grain size distribution. 
This can affect attempts to interpret polarisation behaviour around jet breaks (e.g. \citealp{Lazzati021004}) or models using the absolute level of polarisation (e.g. Gruzinov \& Waxman 1999; Gruzinov 1999). For example, the zero polarisation seen in e.g. {\em lin1} could in principle be an unfortunate effect of HGIP on an intrinsically non-zero polarised afterglow.

We can utilize the reasonable assumption that over the optical range, the intrinsic afterglow linear polarisation will be wavelength independent as it is synchotron emission, whereas HGIP will be strongly wavelength dependent. In the Galactic ISM, the polarisation broadly follows an (empirical) relation known as the Serkowski law:
\[
P_{\rm lin} = P_{\rm lin,max} \exp \left(K \ln^2 \left(\frac{\lambda_{\rm max}}{\lambda} \right)\right).
\]
In this relation, $P_{\rm lin,max}$ is the maximum induced linear polarisation at wavelength $\lambda_{\rm max}$.  The wavelength $\lambda_{\rm max}$ traces the size (distribution) of the dust grains responsible for the observed polarisation, and as such is closely linked to $R_V = A_V / E(B-V)$. \cite{klose} considered the effect of redshift in the Serkowski law, which results in a substitution
$\lambda_{\rm max} \rightarrow (1+z)\lambda^{\rm host}_{\rm max}$ in the equation above, and the relation $\lambda^{\rm host}_{\rm max} = R_V/5.5$. \cite{klose} derive the observed linear
polarisation $P_{\rm lin}/P_{\rm lin,max}$ in standard photometric broadband filters of an intrinsically unpolarised afterglow due to the presence of dust in the host, as a function of redshift $z$ and $R_V$.
Using the results in \cite{klose}, we generate predictions for observed $P_{\rm lin}/P_{\rm lin,max}$ in several broadband filters as a function of $R_V$, using $z=0.971$, shown in Figure \ref{fig_ppmax}.
From these, a predicted ratio of $P_{\rm lin, R} / P_{\rm lin, K}$ is produced as a function of $R_V$, plotted in the lower panel of Figure \ref{fig_ppmax}. 
Multi-wavelength linear polarimetry has been possible for only a small number bursts: 020813 (\citealp{Barth,Lazzati020813}), 021004 (\citealp{Lazzati021004}) and 030329 (\citealp{klose,Greiner}). In these cases, no significant evidence for HGIP has been observed. 

Our $Ks$ band polarimetry is approximately simultaneous to the $R$ band polarimetry epoch {\em lin9} (to be exact, {\em lin9} is simultaneous with {\em linK3}). The (GIP corrected) Stokes parameters for {\em lin9} and {\em linK}
are very similar, though we caution that the data were taken during what appears to be a bump in the polarisation light curve (Fig. \ref{fig_linnight1}). 
The resulting position angles are identical within errors, and we find $P_{\rm lin, R} / P_{\rm lin, K} =  0.72 \pm 0.45$.  This, combined with the strong detected variation of both polarisation degree and angle, demonstrates that (within errors) there is no indication of substantial HGIP contribution to the observed polarisation. \cite{Martinangel} demonstrate that the ISM induces a small degree of circular polarisation even in unpolarised sources. However, the levels of induced circular polarisation are an order of magnitude or more below the limit we set in Section \ref{sec:circpol}, considering the measured Galactic GIP and the low extinction in the host galaxy (Section \ref{sec:lc}).

In the following we therefore proceed from the assumption that the GIP-corrected polarisation is all intrinsic to the source. 
In a future paper (Paper 2; Wiersema et al. 2012 in prep.) we will further quantify the implication of this measurement (and those of other GRB afterglows) on the dust and gas properties in GRB sightlines. A possibility that we will also address in Paper 2 is the effect of dust destruction by the GRB and afterglow high energy radiation, which may make the effect of HGIP time-dependent.

\subsection{Linear polarisation light curve}
Our $R$ band linear polarimetric monitoring took place over three observing nights, as described in Section \ref{sec:linacq}. The behaviour over the first two nights is illustrated in
Figures \ref{fig_linnight1} and \ref{fig_linnight2}, shown in linear time coordinate. On the third night only one deep, final, measurement was taken.

In the first night (see Figure \ref{fig_linnight1}), the source starts off with very low (consistent with zero, within the errors dominated by the uncertainty in the GIP Stokes parameters) polarisation, the lowest observed for a GRB afterglow to date. Following that, a broad, low amplitude bump may be present in the polarimetric curve. GRBs 030329 and 021004 showed a correlation between linear polarisation behaviour and light curve bumps (e.g. \citealp{Greiner}). In Figure \ref{fig_linnight1} we indicate the times of the three bumps we find in our data. Some variability may be present related to these bumps, but the statistical significance is very low. 
After the first eight datapoints obtained at the start of the night, observations were resumed at the end of the night, showing a decline from a higher polarisation level (Table 2), confirmed by the ISAAC polarimetry taken quasi-simultaneous to {\em lin9}. Over the first night, some low level variation in position angle may be present, but the uncertainties are large because of the low linear polarisation: overall, the first night data are consistent with a constant polarisation angle. 

The second night data (see Figure \ref{fig_linnight2}) displays short time scale variability: a polarisation bump, peaking at $\gtrsim3\%$ polarisation, accompanied by a clear rotation of the position angle. Following this, the end of the night shows a slow rise in polarisation, accompanied by further position angle variation. By the third night, the afterglow had become too faint for detailed sampling, and one deep datapoint, {\em lin20}, was obtained (Table 2). This last datapoint clearly shows low-level contribution from the host galaxy in the received light.
We will attempt to interpret the polarimetric data in Section \ref{sec:discussion}.

\subsection{Environment of the burst}
In a forthcoming paper on this dataset (Paper 2) we will analyse the GRB environment in detail. In this section we will derive some properties most relevant to the afterglow physics, and to show that this burst displays no indications of being unusual.  

\begin{figure}   
\includegraphics[width=8.5cm]{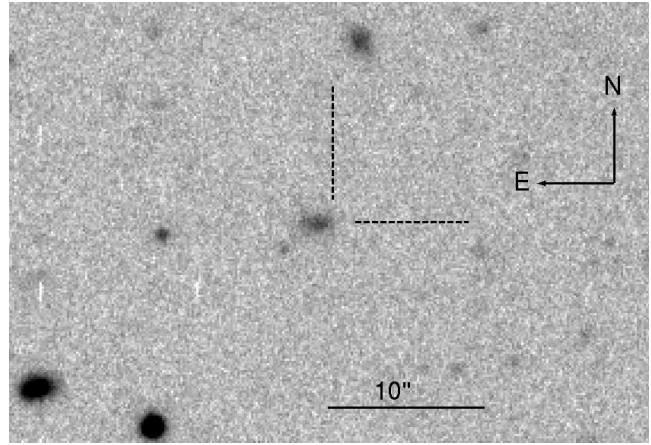}
\caption{The host galaxy of GRB\,091018, from the second epoch Gemini GMOS $r$ band data. The dashed tickmarks mark the position of the GRB as determined from PSF-matched image subtraction of Gemini epochs 1 and 2.}
\label{fig:host}
\end{figure}

In the late time images taken with Gemini (see Table \ref{table:photlog}) the host galaxy can clearly be seen as an extended structure, see Figure \ref{fig:host}, 
visibly extended in East-West direction. An isophote fit using elliptical isophotes (using the {\it ellipse} package within IRAF) gave a best fit for an ellipse orientation
of $89\pm4$ degrees. We determined the position of the GRB within its host by performing image subtraction of the three Gemini epochs using a modified version of the
ISIS2 code (Alard \& Lupton 1998). We detected a clear afterglow residual in a subtraction of the first and third epoch, and between the first and second epochs. No residual was detected in a subtraction between the second and third epochs. We found the GRB location to be offset by approximately 5.8 kiloparsec from the centre of the galaxy (as found from the elliptical isophote fits), see Fig \ref{fig:host}. This offset, as well as the observed host magnitude, $r = 23.4$ (AB; see Table \ref{table:photlog}), is well within the observed distribution for long GRBs at this redshift (e.g. \citealp{Bloom,savaglio}).

\begin{figure*}   
\includegraphics[width=15cm]{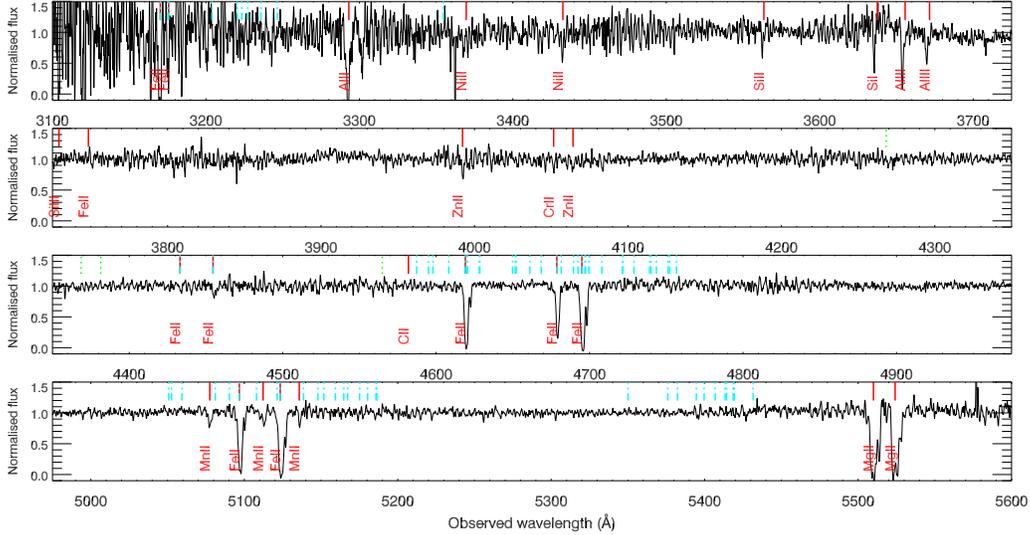}
\caption{The UVB arm spectrum from X-shooter, smoothed with a 3 pixel boxcar for presentation purposes. Some commonly observed (resonance) absorption lines are 
indicated with red, solid, vertical lines and are labelled. The dash-dotted, blue, vertical lines mark the positions of excited finestructure transitions of Fe\,{\sc II}, which are occasionally observed as highly variable lines in GRB afterglow spectra. The dotted, green, vertical lines mark the positions of excited finestructure transitions of Ni\,{\sc II}. 
}
\label{fig:xshspectra}
\end{figure*}

The X-shooter afterglow spectra show very strong absorption lines (Fig \ref{fig:xshspectra}) composed of at least three discrete velocity components, spanning $\sim300$ km s$^{-1}$.  
No lines from intervening systems are detected.
At the GRB redshift we detect both common resonance lines and a small number of transitions arising from excited finestructure and metastable levels (from Fe\,{\sc II} and possibly Ni\,{\sc II}), though the lines from these excited transitions are very weak in our spectrum.  
Time resolved spectroscopy of GRB afterglows has shown that these excited transitions are excited through indirect UV-pumping (fluorescence) by the afterglow photons (e.g. \citealp{finestr}). This unambiguously identifies the redshift  $z = 0.971$ to be that of the GRB.

The X-shooter spectra were taken with the slit position angle aligned with the parallactic angle, set at -90.8$^{\circ}$ at the start of observations, and therefore
oriented broadly along the host galaxy. In addition to absorption lines, we therefore also detect nebular emission lines in the spectrum (e.g. H\,$\alpha$,  [O\,{\sc II}],  [O\,{\sc III}]), though their signal to noise is low as the afterglow strongly dominates the light (see Table \ref{table:photlog}). 
The detection of the resolved [O\,{\sc II}] $\lambda\lambda\,3726,3729$ doublet is shown in 
Figure \ref{fig:o2}. Its flux of $\sim6 \times 10^{-17}$ erg\,s$^{-1}$\,cm$^{-2}$ can be used to derive an estimate of the starformation rate: using the equation from \cite{Kennicutt} we find a SFR\,([O\,{\sc II}]) $\sim4$ Solar masses per year. If we instead use the empirical conversion from [O\,{\sc II}] line luminosity to star formation rate derived 
by \cite{savaglio} for GRB host galaxies, we find SFR\,([O\,{\sc II}]) $\sim1.6$ Solar masses per year (note that no extinction correction was applied to the [O\,{\sc II}] line luminosity).  This SFR\,([O\,{\sc II}]) is fairly typical for long GRB hosts (\citealp{savaglio}). 
In conclusion, the afterglow spectrum, host magnitude and morphology, and star formation rate are all in line with expectations for a normal long GRB host galaxy. 

In Paper 2 we will further analyse the spectrum, and exploit the rare opportunity where we can probe (interstellar) gas and dust in the host galaxy using five methods simultaneously: through afterglow absorption lines (including excited finestructure lines), nebular emission lines, afterglow SEDs, wavelength dependent linear polarimetry and host imaging. 

\section{Discussion}\label{sec:discussion}
\subsection{GRB 091018 compared to GRBs 030329, 021004, 020813, 020405 and 090102}
It is instructive to compare the properties of the polarisation of the afterglow of GRB\,091018 with those of the best studied previous cases: the pre-{\em Swift} bursts 030329 ($z=0.1687$), 021004 ($z=2.323$), 020813 ($z=1.254$) and 020405 ($z=0.690$). Each has multiple measurements over more than one night. We also compare to {\em Swift} burst 090102 ($z=1.547$), which has only one single measurement, which probes the poorly explored early-time afterglow. 

Both 021004 and 030329 have well studied bumps in their afterglow light curves (021004: 
\citealp{Bersier021004,Fox021004,Holland021004,Mirabal021004,Pandey021004}. 030329: \citealp{Bloom030329,Burenin030329,Lipkin030329,Matheson030329,Sato030329,Uemura030329}), frequently attributed to late-time energy injection (late time activity of the central engine).  
Both sources are pre-{\em Swift} bursts and therefore have relatively poorly sampled X-ray light curves compared to bursts in the {\em Swift} era, but well sampled optical and radio afterglows. 
GRB\,030329 has the best sampled polarimetric light curve of any burst to date, with polarisation measurements extending to a month after burst (\citealp{Greiner}).
The data show some correlation between the degree of linear polarisation $P_{\rm lin}$ and changes in the decay index of the afterglow (bumps), though $P_{\rm lin}$ and the light curve do not change strictly simultaneously (\citealp{Greiner}). Similarly, it appears that not every bump is associated with a polarisation change.  Variability in both position angle and polarisation degree is detected for 021004 (e.g. \citealp{Rol,Lazzati021004}), but an association with flare behaviour is not easy to make, in part because the polarimetric coverage is less dense than for 030329. In contrast to 021004 and 030329, the afterglow of 020813 is smooth, and therefore this afterglow may be better suited to study global jet properties. Similarly to 021004 and 030329, variability in polarisation is detected in 020813. 
GRB\,091018 is the first {\em Swift} burst to join the small group of afterglows with observed polarimetric variability. Bumps are observed in the broadband optical light curves of this source, but their amplitude is considerably less than those seen in the afterglows of GRBs 021004 and 030329. If the bump amplitude correlates with polarisation amplitude, we may expect the bumps to have a smaller effect on the observed overall polarisation, and be in a better position to probe the properties of the jet as a whole.

GRB\,020405 is of interest mainly because of the short time scale variability with high polarisation amplitude (nearly 10\% polarisation) that this source appears to
show (\citealp{Covino,Bersier}), though near simultaneous measurements described in \cite{Masetti} do not confirm this behaviour. No jet break was seen in the afterglow light curve, and the polarisation curve could not be reconciled with standard models.

Jetted fireball models predict a clear change in polarisation properties around the jet break time. The behaviour of $P_{\rm lin}$ and the polarisation position angle depends sensitively on the angle between the jet axis and the observer, the internal structure of the jet, the magnetic field properties within the jet and whether or not sideways spreading of the jet occurs at around the jet break time (e.g. \citealp{Rossi} and references therein). Attempts to fit these models to afterglow polarisation data have been a little disappointing: in 030329 and 021004 the bumps in the light curve may have influenced the fits (see e.g. \citealp{Lazzati021004}), whereas the polarimetric sampling of 020813 was rather too low to favour a single jet model, though several models could be excluded (\citealp{Lazzati020813}). The afterglow of 020405 unfortunately shows no jet break within the interval of polarimetric monitoring.   
The very well sampled 030329 has the relative disadvantage of a bright supernova contributing at late times, in addition to the bumps at early times, and determining a unique jet break time proved complicated (with best model likely requiring two jet breaks, see e.g. \citealp{ali030329} and references therein). Our data of 091018 may therefore be one of the first cases where a jet break can be probed in some detail through polarimetry, if the light curve break discussed in Section \ref{sec:lc} is indeed a jet break.

Our polarimetric followup starts at restframe time 0.067 days (5782 seconds) after burst trigger, considerably earlier than the earliest datapoints of 030329, 021004 and 020813.
It is clear that polarimetry very early after burst requires automated/robotic procedures. Using such an approach, a very early afterglow polarimetric detection of GRB\,090102 was obtained by \cite{Steele}, who secure a detection of $P_{\rm lin} = 10.2\pm 1.3 \%$ at restframe time 75 seconds after
burst, using the robotic Liverpool Telescope. Such a high degree of polarisation, measured when the reverse shock component was still bright, is consistent with the presence of large-scale ordered magnetic fields. Although our data on 091018 are taken considerably later in time than those on 090102, the superb sensitivity of the VLT may allow us to continue to probe the magnetic field structure at late time through our high-precision linear and circular polarimetry.

\subsection{GRB 091018 and models}
\subsubsection{Reverse shock}
The start of our polarimetric monitoring was sufficiently late that a contribution of reverse shock emission to the received light is likely negligible.  While this precludes
study of some magnetisation properties (e.g. \citealp{radiocirc}), this makes a comparison of the data with models easier, as only a forward shock model needs to be considered. 
The low linear polarisation at the start of our monitoring appears to confirm this: emission from reverse shocks is expected to be considerably polarised, as the shock travels back into magnetized shocked material (e.g. \citealp{radiocirc,Steele}). 

\subsubsection{Jet break}
We compare our data with the model series computed by \cite{Rossi}, hereafter referred to as R04, who derive the expected linear polarisation curves (including position angle) for a grid of different jet model parameters, assuming there is no additional coherent component of the magnetic field present, and assuming HGIP is negligible. In particular, R04 consider the cases of homogeneous jets, structured jets and Gaussian jets, each with a grid of different viewing angles (in the case of homogenous jets this is the angle the line of sight makes with the jet axis), and for a range of wavelengths and physical parameters. They further distinguish jet breaks with and without sideways spreading. 

We can briefly summarize the main differences between the various models as found by R04, and refer for details and explanations to that paper (see also Figure 18 of R04):\\
{\em (i)}: Homogeneous jets show two peaks in the polarisation curves, with the second peak always reaching a higher polarisation than the first. Structured and Gaussian jets show one peak only, with
peak time for a Gaussian jet significantly later than for a structured jet, which has its peak around the time of the break. The peaks of the structured and Gaussian jets are much wider than the ones for homogeneous jets (i.e. rise time and decay time scales are long);\\
{\em (ii)}: The position angle for a homogeneous jet changes by 90$^{\circ}$ in between the two peaks in the polarisation curves: the two peaks each have constant position angle, but are 90$^{\circ}$ different. Structured and Gaussian jets have constant position angles throughout;\\
{\em (iii)}:  Structured jets show non-negligible polarisation at early times, whereas homogenous jets and Gaussian jets have zero polarisation at early times;\\
{\em (iv)}: The shape of the bumps in the different jet structures is different;\\
{\em (v)}: Sideways expanding jets show lower polarisation values than non sideways expanding jets (all other parameters being equal), with wider bump(s) in the polarisation curve.

\begin{figure}   
\includegraphics[width=8.5cm]{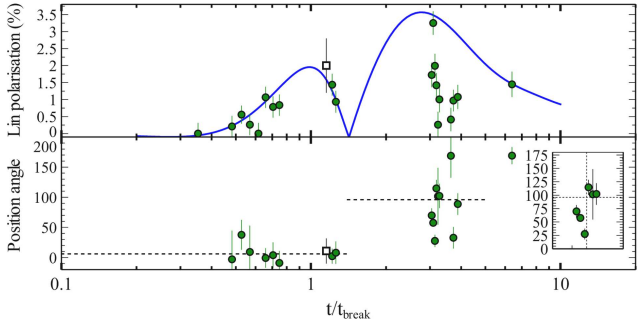}
\caption{All GIP-corrected linear polarisation data plotted as a function of the time since burst divided by the time of the break in the optical 
light curve (i.e. $t_{\rm break} = 3.23 \times 10^4$ seconds). Overplotted in blue is the model (from 
\citealp{Rossi}) for a homogeneous jet with a jet break at $t_{break}$ that does not experience sideways expansion, with viewing angle $\theta_{\rm obs} =0.2\theta_{\rm jet}$. The model has been shifted
slightly (by dividing the model $t/t_{\rm jet}$ by 1.15). The lower panel shows the polarisation position angle. The dotted line for $t/t_{\rm jet}< 1.5$ shows the average position angle (6 degrees). The dotted line for $t/t_{\rm jet}>1.5$ is set at 96 degrees: the average value before this time plus 90 degrees, to show the predicted 90 degree position angle swing. The inset shows the rapid polarisation rotation at $t/t_{\rm break}\sim3.16$ (vertical dotted line in inset), with the horizontal dotted line at a position angle of 96 degrees.
}
\label{fig:polanorm}
\end{figure}

We start by evaluating the data in time coordinates normalised by the time of the break seen in the optical data (see Section \ref{sec:lc}). It is clear that some polarisation behaviour is observed occurring close to $t/t_{\rm break} \sim 1$, as expected for jet breaks.
 In the case that this break is a jet break, we can then directly compare the curves with the ones in R04. Fig.~\ref{fig:polanorm} shows a plot in these units. Comparing to the points {\em i - v} listed above, there appears to be evidence that: a model with constant position angle throughout will not provide a good fit; early time polarisation is consistent with zero (or at least is lower than predictions for structured jets); a single bump model with a shape as predicted by the models in R04 seems inconsistent with the data. 
On Fig.~\ref{fig:polanorm} we therefore plot a model for a homogeneous, non-sideways expanding jet with viewing angle $\theta_{\rm obs} =0.2\theta_{\rm jet}$ (from Figure 8 in R04). We shift this curve slightly in time coordinate by dividing $t/t_{\rm jet}$ by 1.15 to get a better visual match. The first bump in the polarisation curve appears to be well reproduced in both shape
and amplitude, including the zero polarisation at early times. This model predicts a constant polarisation position angle over the first bump. We indicate in Fig.~\ref{fig:polanorm} (dotted line) the average of the position angle at $t/t_{\rm break} \lesssim1.4$. As required by the R04 model, the position angle is constant over this bump. 

The second bump  predicted by the homogeneous jet model should have a larger amplitude than the first, a position angle 90$^\circ$ offset from the first bump, and a morphology opposite to the first (in coordinates used in Fig \ref{fig:polanorm}).  Some of the data points indeed have a higher polarisation than seen in the first bump, but the morphology of the bump as predicted by the model is not observed. We examine the behaviour of the position angle by adding 90$^{\circ}$ to the position angle for $t/t_{\rm break} \lesssim1.4$. In Fig.~\ref{fig:polanorm} this is the second horizontal dotted line. It is clearly seen that the value of the position angle is overall higher than in the first bump, but is not constant as the model predicts, showing considerable variability on a short time scale (possibly varying around the predicted position angle), including a rotation of the position angle visible in the interval  $t/t_{\rm break} \sim3.0 - 3.3$. The polarisation curve shows, compared to the homogeneous jet model, a deep short dip, in the interval $t/t_{\rm break} \sim3 - 4$. This dip shows a very steep decay in the first few datapoints, simultaneous with the observed rotation in position angle. This behaviour is not predicted by any jet break model. 

We could consider the possibility that there is an additional polarisation curve component (hereafter simply referred to as the extra component) on top of a slower behaviour as predicted by jet break models. A third smooth, fairly narrow, polarised bump peaking at around $t/t_{\rm break}\sim3-4$ could be present, having a position angle significantly offset from the $\theta = 96^\circ$ of the second bump from the homogenous jet model. The sum of this component and that of the (second bump of the homogenous) jet model can give rise to most of the observations: the steep dip, where the contributions of the two components nearly cancel due to differing position angles, and the rotation in position angle, where the components are at similar amplitude but very different position angle. 
However, this is a rather degenerate problem, as the amplitude as a function of time (the shape) of the extra component and the position angle as a function of time are all unknown. A component with a fixed position angle but variable amplitude (i.e. a component much like the polarisation peaks from a jet break) can approximate the data (with a position angle $\sim70$ degrees offset from the constant $96^{\circ}$, and peaking around $t/t_{\rm break}\sim3.5$), but a component with nearly constant amplitude but position angle varying on very short time scales can provide a reasonable approximation too.  \cite{doublejet} evaluate the expected polarisation curves of
afterglows from two-component jets (as proposed for e.g. 030329, \citealp{Berger030329}), which show, as expected, a large variety of behaviour, depending mostly on viewing angle and contrast between the narrow and wide jet components. However, the observed variability in the data of 091018 around $t/t_{\rm break} \sim 3 - 4$ is much more rapid than 
predicted by the models of \cite{doublejet}.
At the current data quality, and including the uncertainty in and applicability of the underlying model (homogeneous jet model from R04), we are not in a position to quantify
the behaviour of this putative component further. As the shape of the bump and time variability of the position angle are unclear, it is difficult to speculate on the possible origin of such an additional component. The feature occurs after the expected time of the peak of the second bump in the R04 model and no such feature is visible at $t < t_{\rm break}$. As such, it may be related to a discrete bright region (patch) on the jet that comes into view just after the second bump peaks. In this scenario we may expect a constant position angle of the component produced by the patch (if the field in the patch is coherent), with a bump shape reflecting the brightness profile of the bright patch. However, other explanations, such as microlensing, two-component jets or others, are difficult to rule out with the polarisation data alone.

The time span possibly most affected by the extra component, $t/t_{\rm break} \sim3 - 4$, corresponds to 
$9.7 - 12.9 \times 10^4$ seconds. In the optical light curves no significant slow variability at around that time is obvious. The XRT data show a possible broad, slow softening around this time, but our SED fits do not point to strong spectral variation at this time. The shallow post-break decay indices and/or the flares observed in the optical light curve (Section
\ref{sec:lc}) may be related to the observed polarisation behaviour, i.e. be linked to the observed extra component, but further modelling will be required to test this. 

There is at least one other source that shows a variability in the polarisation curve on short time scales (intranight) and unrelated to a jet break, GRB\,020405. However, in that source only a single datapoint showed the deviation from a constant polarisation degree (\citealp{Bersier}), and the position angle was not seen to vary over the available polarisation measurements. No features resembling jet break bumps were detected in that source, consistent with the absence of a jet break in the light curves. The possible detection of a variation at fairly short time scales, with possibly a constant position angle and likely unrelated to a jet break in GRB\,091018 (the extra component in the polarisation curve) may be an indication that this type of behaviour is indeed real.

The conclusion from the above discussion is that none of the simple models of polarisation behaviour of GRBs directly reproduces the observed behaviour. Of the available models, the homogeneous jet break model comes closest, but requires at least one additional component, the nature of which is currently unknown. 

\subsubsection{Ordered fields}
The circular polarisation upper limit derived in Section \ref{sec:circpol} can be used to probe large scale ordered fields, which have been proposed by several authors to explain various phenomena, such as the high linear polarisation at very early times (\citealp{Steele}), or the lack of obvious polarisation position angle shifts around the jet break time for e.g. GRBs 030329 and 021004 (\citealp{Granot}). 
Models of the expected circular polarisation in optical wavelengths for an external shock (\citealp{masu,sagiv,Toma}) generally find that afterglows should show very low circular polarisation in the absence of strong large scale ordered fields.  Our detection limit of $P_{\rm circ} < 0.15\%$ is in full agreement with these predictions 
(see e.g. \citealp{Toma}), and excludes strong ordered field components.
 We note the timing of our circular polarimetry: our data were taken when the reverse shock emission is likely negligible, and when the linear polarisation was still very weak. 
It is clear that it is not easy to significantly improve on the depth of this limit in optical wavelengths, but it would be interesting to search for circular polarisation in other bursts, to establish whether this is a common feature, and to search for circular polarisation at early times when reverse shock emission is still present.

\subsubsection{Energy injection}
An alternative explanation of the observed break in the light curves is that this break marks the end of (continuous) energy injection, though the post-break decay indices are not as expected from the closure relations (Section \ref{sec:lc}). The occurrence of significant polarimetric variability at around $t/t_{\rm break} \sim1$ indicates that if this scenario is correct, the end of energy injection must be accompanied by a geometric change, as the polarisation position angle stays constant. In a study of GRB\,021004, \cite{Bjornsson} model the 
strong rebrightenings visible in the afterglow light curve with discrete, powerful, energy injections, and demonstrate that the light curves are fairly well reproduced this way. They demonstrate that their model shows some correlation with the polarimetric data, i.e. that polarisation is related to bumps. For GRB\,030329, \cite{Greiner} come to a similar conclusion from their data. We see at least three low amplitude bumps in the optical light curve, at times given in Section \ref{sec:lc}, but a correlation with polarimetric behaviour is not clearly visible, largely because of the larger uncertainties in our polarisation values compared to the much brighter afterglow of 030329. Similarly, a possible relation between the extra component in the polarisation curve and the flaring around the jet break time is unclear. Future (numerical) models of 
light curve bumps/flares (e.g. \citealp{Vlasis}) and high quality data (e.g. \citealp{kruehler}) may provide the possibilities to compute expected polarisation signals from the simulations. These may then be used to understand the connection between the physical parameters behind these bumps and the behaviour in polarised light.

\section{Conclusions}\label{sec:conclusions}
We presented a large polarimetry dataset of the afterglow of GRB\,091018, and add to that an extensive multicolour photometric dataset and medium resolution spectroscopy.
We find an abundance of absorption lines in the spectrum of the afterglow as well as a few emission lines, from which we find a redshift $z=0.971$.  
The  X-ray light curve shows at least two breaks, as well as a complex of flaring activity. The densely sampled optical light curve shows several flares as well, and through a fit using a broken power law and three Gaussian components (to describe the flares), we find an achromatic break at $t_{\rm break} \sim3.2 \times 10^4$ seconds. Our polarisation data cover this break, allowing us to probe its nature using linear polarimetry.
Our polarimetry results can be summarised as:
\begin{itemize}
\item In contrast to GRBs 030329 and 021004, there are no strong rebrightenings that confuse our view of the polarisation behaviour.
\item Our deep circular polarimetry at $t\sim0.15$ days after burst shows no circular polarisation, with $2\sigma$ limit $P_{\rm circ} < 0.15\%$.
\item The quasi-simultaneous $R$ and $Ks$ band polarimetry shows no strong indications of a considerable contribution of polarisation from the ISM in the host galaxy.
\item The linear polarimetry shows a constant position angle for $t/t_{\rm break} < 1.4$, and a considerably higher, but strongly variable, value thereafter.
\item At least two peaks are present in the polarisation curve. The first of these, as well as the constant position angle,  is well reproduced by a model of a jet break using a homogenous, non-sideways spreading jet with small viewing angle (from R04).   
\item If the observed break is a jet break, and if the model of a homogenous, non-sideways spreading jet is correct, an additional component is required to reproduce the observed polarisation curve and position angles. This may be a short duration bump with constant position angle, offset from the position angle of the underlying jet break component. 
\end{itemize}

From the above it is clear that the models for polarisation behaviour around the time of jet break (as in R04) can not, in their current form and on their own, explain the full behaviour of the polarisation curve of 091018. Further modelling efforts, in particular into the effects of rebrightenings / bumps and continuous energy injection on the observed polarisation, are vital to exploit the full potential of polarimetry as probe of afterglow physics. 

These results are the first of their kind for {\em Swift} GRBs, and indeed GRB\,091018 can be considered a ``normal'' GRB, with none of its prompt emission or afterglow properties particularly remarkable, in contrast to some of the polarimetrically studied pre-{\em Swift} bursts like 030329. This is in part because our selection criteria of sources to study with polarimetry were only that: the source had to be a {\em Swift} burst with a detected afterglow in both XRT and UVOT data. This avoids a bias towards the brightest or slowest decaying optical afterglows, and indeed it is clear from Section \ref{sec:lc} that GRB\,091018 is one of the faintest afterglows with multi-night polarimetry.  Further, similar, observations of ({\em Swift}) bursts with different micro- and macro-physical parameters (in particular energetics, jet opening angle, density $n$ and density gradient $k$) will be required to establish whether or not GRB\,091018 is representative of the sample of GRB afterglows as a whole. 

Unfortunately, no long wavelength
(radio, (sub-)mm) observations were available for GRB\,091018. A deep polarimetric campaign on an afterglow, with radio and (sub-)mm light curve data in addition to optical and X-ray light curves would be particularly interesting, since with the resulting broadband SEDs (with sensitivity redwards of the synchrotron peak frequency $\nu_m$) we 
would be able to fit for parameters that are particularly relevant for polarimetric interpretations, e.g. $\epsilon_E$ and $\epsilon_B$.

\section*{Acknowledgments}
We warmly thank the ESO staff for obtaining the VLT data discussed here. We thank Justyn Maund, Dave Russell, Daniele Malesani, Piergiorgio Casella, Elena Rossi, 
Christina Th\"{o}ne and Alexander Kann for their help and useful discussions. We thank the anonymous referee for useful comments and suggestions.
Based on observations made with ESO Telescopes at the Paranal Observatory under programmes 084.D-0949 and 084.A-0260. 
KW acknowledges support from STFC. TK acknowledges support by the European Commission under the Marie Curie Intra-European Fellowship Programme. The Dark Cosmology Centre is funded by the Danish National Research Foundation. RLCS is supported by a Royal Society Fellowship.
 The financial support of the British Council and Platform Beta Techniek through the Partnership Programme in Science (PPS WS 005, PI Wiersema) is gratefully acknowledged.
Part of the funding for GROND (both hardware as well as personnel) was generously granted from the Leibniz-Prize to Prof. G. Hasinger (DFG grant HA 1850/28-1).                                                                                          
SK and AR acknowledge support by DFG grant Kl 766/16-1, and AR in addition from the BLANCEFLOR Boncompagni-Ludovisi, n\'ee Bildt foundation.
This work made use of data supplied by the UK {\em Swift} Science Data Centre at the University of Leicester.
IRAF is distributed by the National Optical Astronomy Observatory, which is operated by the Association of Universities for Research in Astronomy (AURA) under cooperative agreement with the National Science Foundation.

\begin{table}
 \centering
 \scriptsize
  \caption{Log of our photometry.  All magnitudes are AB magnitudes. These magnitudes are not corrected for the Galactic foreground extinction. $^{\rm (a)}$: calibrated onto Sloan $r$ using GROND.
  }\label{table:photlog}
  \begin{tabular}{@{}llll@{}}
  \hline
  Inst. / Filter                     &    Time since burst   &  Exp. time                       & Magnitude        \\ 
                                          &      (days)                    & (seconds)                    &                     \\
 \hline
 GROND, $g$ &0.12693	&115	&$19.06		\pm0.04$\\
&0.12906	&115	&$19.05		\pm0.03$\\
&0.13125	&115	&$19.08		\pm0.02$\\
&0.13344	&115	&$19.06		\pm0.02$\\
&0.13575	&115	&$19.06		\pm0.02$\\
&0.13793	&115	&$19.07		\pm0.02$\\
&0.14012	&115	&$19.08		\pm0.01$\\
&0.14231	&115	&$19.09		\pm0.01$\\
&0.14457	&115	&$19.11		\pm0.02$\\
&0.14670	&115	&$19.12		\pm0.02$\\
&0.14889	&115	&$19.14		\pm0.02$\\
&0.15109	&115	&$19.12		\pm0.02$\\
&0.15330	&115	&$19.15		\pm0.02$\\
&0.15551	&115	&$19.15		\pm0.02$\\
&0.15762	&115	&$19.14		\pm0.02$\\
&0.15985	&115	&$19.15		\pm0.02$\\
&0.16216	&115	&$19.16		\pm0.02$\\
&0.16429	&115	&$19.17		\pm0.02$\\
&0.16648	&115	&$19.17		\pm0.02$\\
&0.16867	&115	&$19.16		\pm0.02$\\
&0.17096	&115	&$19.17		\pm0.02$\\
&0.17318	&115	&$19.22		\pm0.02$\\
&0.17541	&115	&$19.23		\pm0.02$\\
&0.17765	&115	&$19.26		\pm0.02$\\
&0.17998	&115	&$19.32		\pm0.02$\\
&0.18214	&115	&$19.33		\pm0.02$\\
&0.18437	&115	&$19.34		\pm0.02$\\
&0.18660	&115	&$19.34		\pm0.02$\\
&0.18891	&115	&$19.35		\pm0.02$\\
&0.19115	&115	&$19.34		\pm0.02$\\
&0.19333	&115	&$19.36		\pm0.02$\\
&0.19552	&115	&$19.38		\pm0.02$\\
&0.19781	&115	&$19.39		\pm0.02$\\
&0.19996	&115	&$19.42		\pm0.02$\\
&0.20214	&115	&$19.46		\pm0.02$\\
&0.20433	&115	&$19.47		\pm0.02$\\
&0.20667	&115	&$19.48		\pm0.02$\\
&0.20886	&115	&$19.48		\pm0.02$\\
&0.21105	&115	&$19.51		\pm0.02$\\
&0.21324	&115	&$19.54		\pm0.02$\\
&0.21556	&115	&$19.54		\pm0.02$\\
&0.21767	&115	&$19.53		\pm0.02$\\
&0.21985	&115	&$19.54		\pm0.02$\\
&0.22204	&115	&$19.54		\pm0.02$\\
&0.22433	&115	&$19.53		\pm0.02$\\
&0.22648	&115	&$19.51		\pm0.02$\\
&0.22867	&115	&$19.52		\pm0.02$\\
&0.23086	&115	&$19.54		\pm0.02$\\
&0.23313	&115	&$19.54		\pm0.02$\\
&0.23526	&115	&$19.54		\pm0.02$\\
&0.23749	&115	&$19.54		\pm0.02$\\
&0.23977	&115	&$19.54		\pm0.02$\\
&0.34860	&675	&$20.01		\pm0.02$\\
&0.44872	&688	&$20.21		\pm0.02$\\
&1.20917	&675	&$21.76		\pm0.04$\\
&1.30936	&688	&$21.82		\pm0.04$\\
&1.49485	&688	&$22.02		\pm0.05$\\
&2.22673	&1727	&$22.53		\pm0.04$\\
&3.15788	&1722	&$22.94	\pm0.07$\\
 \hline
 GROND, $r$ &0.12693	&115	&$18.87	\pm0.03$\\
&0.12906	&115	&$18.86	\pm0.02$\\
&0.13125	&115	&$18.85	\pm0.02$\\
&0.13344	&115	&$18.87	\pm0.02$\\
&0.13575	&115	&$18.88	\pm0.02$\\
&0.13793	&115	&$18.91	\pm0.02$\\
&0.14012	&115	&$18.91	\pm0.02$\\
&0.14231	&115	&$18.92	\pm0.02$\\
&0.14457	&115	&$18.93	\pm0.02$\\
&0.14670	&115	&$18.93	\pm0.02$\\
&0.14889	&115	&$18.94	\pm0.02$\\
&0.15109	&115	&$18.94	\pm0.02$\\
&0.15330	&115	&$18.94	\pm0.02$\\
&0.15551	&115	&$18.96	\pm0.02$\\
&0.15762	&115	&$18.97	\pm0.02$\\
&0.15985	&115	&$18.98	\pm0.02$\\
&0.16216	&115	&$18.98	\pm0.02$\\
\end{tabular}
\normalsize
\end{table}
\setcounter{table}{2}
\begin{table}
 \centering
 \scriptsize
  \caption{ Continued.}
  \begin{tabular}{@{}llll@{}}
  \hline
&0.16429	&115	&$18.99	\pm0.02$\\
&0.16648	&115	&$18.98	\pm0.02$\\
&0.16867	&115	&$18.97	\pm0.02$\\
&0.17096	&115	&$18.99	\pm0.02$\\
&0.17318	&115	&$19.03	\pm0.02$\\
&0.17541	&115	&$19.06	\pm0.02$\\
&0.17765	&115	&$19.08	\pm0.02$\\
&0.17998	&115	&$19.11	\pm0.02$\\
&0.18214	&115	&$19.14	\pm0.02$\\
&0.18437	&115	&$19.15	\pm0.02$\\
&0.18660	&115	&$19.17	\pm0.02$\\
&0.18891	&115	&$19.15	\pm0.02$\\
&0.19115	&115	&$19.17	\pm0.02$\\
&0.19333	&115	&$19.17	\pm0.02$\\
&0.19552	&115	&$19.18	\pm0.02$\\
&0.19781	&115	&$19.21	\pm0.02$\\
&0.19996	&115	&$19.24	\pm0.02$\\
&0.20214	&115	&$19.25	\pm0.02$\\
&0.20433	&115	&$19.27	\pm0.02$\\
&0.20667	&115	&$19.30	\pm0.02$\\
&0.20886	&115	&$19.32	\pm0.02$\\
&0.21105	&115	&$19.33	\pm0.02$\\
&0.21324	&115	&$19.33	\pm0.02$\\
&0.21556	&115	&$19.37	\pm0.02$\\
&0.21767	&115	&$19.36	\pm0.02$\\
&0.21985	&115	&$19.37	\pm0.02$\\
&0.22204	&115	&$19.36	\pm0.02$\\
&0.22433	&115	&$19.33	\pm0.02$\\
&0.22648	&115	&$19.34	\pm0.02$\\
&0.22867	&115	&$19.35	\pm0.02$\\
&0.23086	&115	&$19.35	\pm0.02$\\
&0.23313	&115	&$19.34	\pm0.02$\\
&0.23526	&115	&$19.35	\pm0.02$\\
&0.23749	&115	&$19.35	\pm0.02$\\
&0.23977	&115	&$19.37	\pm0.02$\\
&0.34860	&675	&$19.82	\pm0.02$\\
&0.44872	&688	&$20.01	\pm0.02$\\
&1.20917	&675	&$21.59	\pm0.03$\\
&1.30936	&688	&$21.68	\pm0.03$\\
&1.49485	&688	&$21.88	\pm0.05$\\
&2.22673	&1727	&$22.34	\pm0.04$\\
&3.15788	&1722	&$22.81	\pm0.10$\\
\hline
 GROND, $i$ &0.12693	&115	&$18.71		\pm0.04$\\
&0.12906	&115	&$18.71		\pm0.03$\\
&0.13125	&115	&$18.72		\pm0.02$\\
&0.13344	&115	&$18.75		\pm0.03$\\
&0.13575	&115	&$18.76		\pm0.03$\\
&0.13793	&115	&$18.79		\pm0.02$\\
&0.14012	&115	&$18.79		\pm0.02$\\
&0.14231	&115	&$18.78		\pm0.02$\\
&0.14457	&115	&$18.81		\pm0.03$\\
&0.14670	&115	&$18.80		\pm0.02$\\
&0.14889	&115	&$18.82		\pm0.02$\\
&0.15109	&115	&$18.80		\pm0.02$\\
&0.15330	&115	&$18.84		\pm0.03$\\
&0.15551	&115	&$18.84		\pm0.02$\\
&0.15762	&115	&$18.84		\pm0.02$\\
&0.15985	&115	&$18.84		\pm0.02$\\
&0.16216	&115	&$18.85		\pm0.02$\\
&0.16429	&115	&$18.84		\pm0.02$\\
&0.16648	&115	&$18.84		\pm0.02$\\
&0.16867	&115	&$18.82		\pm0.02$\\
&0.17096	&115	&$18.84		\pm0.02$\\
&0.17318	&115	&$18.88		\pm0.02$\\
&0.17541	&115	&$18.91		\pm0.02$\\
&0.17765	&115	&$18.95		\pm0.02$\\
&0.17998	&115	&$18.97		\pm0.02$\\
&0.18214	&115	&$19.01		\pm0.02$\\
&0.18437	&115	&$19.03		\pm0.02$\\
&0.18660	&115	&$19.02		\pm0.02$\\
&0.18891	&115	&$19.03		\pm0.02$\\
&0.19115	&115	&$19.02		\pm0.02$\\
&0.19333	&115	&$19.02		\pm0.02$\\
&0.19552	&115	&$19.02		\pm0.02$\\
&0.19781	&115	&$19.07		\pm0.02$\\
&0.19996	&115	&$19.09		\pm0.02$\\
&0.20214	&115	&$19.12		\pm0.02$\\
&0.20433	&115	&$19.13		\pm0.02$\\
&0.20667	&115	&$19.15		\pm0.02$\\
&0.20886	&115	&$19.15		\pm0.02$\\
&0.21105	&115	&$19.16		\pm0.02$\\
\end{tabular}
\normalsize
\end{table}
\setcounter{table}{2}
\begin{table}
 \centering
 \scriptsize
  \caption{ Continued.}
  \begin{tabular}{@{}llll@{}}
  \hline
&0.21324	&115	&$19.18		\pm0.02$\\
&0.21556	&115	&$19.20		\pm0.02$\\
&0.21767	&115	&$19.21		\pm0.02$\\
&0.21985	&115	&$19.21		\pm0.02$\\
&0.22204	&115	&$19.18		\pm0.02$\\
&0.22433	&115	&$19.21		\pm0.02$\\
&0.22648	&115	&$19.20		\pm0.02$\\
&0.22867	&115	&$19.21		\pm0.02$\\
&0.23086	&115	&$19.21		\pm0.02$\\
&0.23313	&115	&$19.24		\pm0.03$\\
&0.23526	&115	&$19.23		\pm0.02$\\
&0.23749	&115	&$19.21		\pm0.03$\\
&0.23977	&115	&$19.21		\pm0.03$\\
&0.34860	&675	&$19.67		\pm0.02$\\
&0.44872	&688	&$19.85		\pm0.02$\\
&1.20917	&675	&$21.47		\pm0.06$\\
&1.30936	&688	&$21.51		\pm0.06$\\
&1.49485	&688	&$21.64		\pm0.10$\\
&2.22673	&1727	&$22.03		\pm0.08$\\
&3.15788	&1722	&$22.53	         \pm0.13$\\
\hline
GROND, $z$ &0.12693	&115	&$18.59	\pm0.05$\\
&0.12906	&115	&$18.58	\pm0.05$\\
&0.13125	&115	&$18.58	\pm0.03$\\
&0.13344	&115	&$18.59	\pm0.03$\\
&0.13575	&115	&$18.61	\pm0.03$\\
&0.13793	&115	&$18.65	\pm0.03$\\
&0.14012	&115	&$18.63	\pm0.03$\\
&0.14231	&115	&$18.65	\pm0.03$\\
&0.14457	&115	&$18.64	\pm0.03$\\
&0.14670	&115	&$18.66	\pm0.03$\\
&0.14889	&115	&$18.70	\pm0.03$\\
&0.15109	&115	&$18.71	\pm0.03$\\
&0.15330	&115	&$18.70	\pm0.03$\\
&0.15551	&115	&$18.68	\pm0.03$\\
&0.15762	&115	&$18.71	\pm0.03$\\
&0.15985	&115	&$18.71	\pm0.03$\\
&0.16216	&115	&$18.68	\pm0.03$\\
&0.16429	&115	&$18.67	\pm0.03$\\
&0.16648	&115	&$18.69	\pm0.03$\\
&0.16867	&115	&$18.68	\pm0.03$\\
&0.17096	&115	&$18.70	\pm0.03$\\
&0.17318	&115	&$18.74	\pm0.03$\\
&0.17541	&115	&$18.80	\pm0.03$\\
&0.17765	&115	&$18.82	\pm0.03$\\
&0.17998	&115	&$18.82	\pm0.03$\\
&0.18214	&115	&$18.87	\pm0.03$\\
&0.18437	&115	&$18.90	\pm0.03$\\
&0.18660	&115	&$18.87	\pm0.03$\\
&0.18891	&115	&$18.89	\pm0.03$\\
&0.19115	&115	&$18.88	\pm0.03$\\
&0.19333	&115	&$18.88	\pm0.03$\\
&0.19552	&115	&$18.89	\pm0.02$\\
&0.19781	&115	&$18.90	\pm0.03$\\
&0.19996	&115	&$18.91	\pm0.03$\\
&0.20214	&115	&$18.94	\pm0.03$\\
&0.20433	&115	&$18.93	\pm0.03$\\
&0.20667	&115	&$18.99	\pm0.03$\\
&0.20886	&115	&$19.03	\pm0.03$\\
&0.21105	&115	&$19.04	\pm0.03$\\
&0.21324	&115	&$19.01	\pm0.03$\\
&0.21556	&115	&$19.06	\pm0.03$\\
&0.21767	&115	&$19.05	\pm0.03$\\
&0.21985	&115	&$19.08	\pm0.03$\\
&0.22204	&115	&$19.09	\pm0.03$\\
&0.22433	&115	&$19.07	\pm0.03$\\
&0.22648	&115	&$19.08	\pm0.03$\\
&0.22867	&115	&$19.10	\pm0.03$\\
&0.23086	&115	&$19.12	\pm0.03$\\
&0.23313	&115	&$19.06	\pm0.03$\\
&0.23526	&115	&$19.05	\pm0.03$\\
&0.23749	&115	&$19.07	\pm0.03$\\
&0.23977	&115	&$19.07	\pm0.03$\\
&0.34860	&675	&$19.51	\pm0.02$\\
&0.44872	&688	&$19.70	\pm0.03$\\
&1.20917	&675	&$21.24	\pm0.07$\\
&1.30936	&688	&$21.32	\pm0.08$\\
&1.49485	&688	&$21.61	\pm0.12$\\
&2.22673	&1727	&$22.00	\pm0.09$\\
&3.15788	&1722	&$22.64	\pm0.20$\\
\hline
\end{tabular}
\normalsize
\end{table}
\setcounter{table}{2}
\begin{table}
 \centering
 \scriptsize
  \caption{ Continued.}
  \begin{tabular}{@{}llll@{}}
  \hline
GROND, $J$&0.13048	&730    &         $	18.34\pm	0.03$\\
&0.13932	&730    &	      $   18.38\pm	0.03$\\
&0.14812	  &730  &  	$18.40\pm	0.03$\\
&0.15688	  &730  &  	$18.45\pm	0.03$\\
&0.16570	  &730  &  	$18.38\pm	0.03$\\
&0.17460	  &730  &  	$18.52\pm	0.03$\\
&0.18359	  &730  &  	$18.58\pm	0.03$\\
&0.19251	  &730  &  	$18.58\pm	0.03$\\
&0.20136	  &730  &  	$18.68\pm	0.03$\\
&0.21024	  &735  &  	$18.73\pm	0.03$\\
&0.21909	 & 727 &   	$18.82\pm	0.03$\\
&0.22789	 & 730 &   	$18.77\pm	0.03$\\
&0.23674	 & 730 &   	$18.79\pm	0.03$\\
&0.34889	 & 730 &   	$19.17\pm	0.03$\\
&0.44903	& 730 &  	$19.37\pm	0.04$\\
\hline
GROND, $H$ & 0.13048	& 730.08&  	$18.19\pm	0.05$\\
& 0.13932	& 730.08&  	$18.18\pm	0.04$\\
& 0.14812	& 730.08&  	$18.24\pm	0.04$\\
& 0.15688	& 730.08&  	$18.27\pm	0.03$\\
& 0.16570	& 730.08&  	$18.22\pm	0.04$\\
& 0.17460	& 730.08&  	$18.31\pm	0.03$\\
& 0.18359	& 730.08&  	$18.34\pm	0.03$\\
& 0.19251	& 730.08&  	$18.38\pm	0.03$\\
& 0.20136	& 730.08&  	$18.51\pm	0.03$\\
& 0.21024	& 734.62&  	$18.53\pm	0.04$\\
& 0.21909	& 726.82&  	$18.60\pm	0.04$\\
& 0.22789	& 730.08&  	$18.55\pm	0.04$\\
& 0.23674	& 730.08&  	$18.57\pm	0.03$\\
& 0.34889	& 730.08&  	$19.04\pm	0.09$\\
&0.44903	& 730.08 &	$19.22\pm	0.09$\\
\hline
GROND, $K$ & 0.13048	& 730 & 	$17.95\pm	0.07$\\
& 0.13932	& 730 & 	$18.06\pm	0.07$\\
& 0.14812	& 730 & 	$17.96\pm	0.06$\\
& 0.15688	         & 730 & 	$18.10\pm	0.06$\\
& 0.16570	         & 730 & 	$18.08\pm	0.06$\\
& 0.17460         & 730 & 	$18.17\pm	0.06$\\
& 0.18359	         & 730 & 	$18.25\pm	0.06$\\
& 0.19251         & 730 & 	$18.21\pm	0.06$\\
& 0.20136	         & 730 & 	$18.37\pm	0.06$\\
& 0.21024	         & 735 & 	$18.32\pm	0.06$\\
& 0.21909	         & 727 & 	$18.35\pm	0.06$\\
& 0.22789	        & 730 & 	$18.32\pm	0.06$\\
& 0.23674	        & 730 & 	$18.42\pm	0.06$\\
& 0.34889	        & 730 & 	$18.83\pm	0.07$\\
& 0.44903	        & 730 & 	$19.07\pm	0.08$\\
 \hline  
FORS2, R$_{\rm special}^{\rm (a)}$   & 0.12384 & 20   &   $18.85  \pm 0.03$\\
                                                   & 0.13707 & 20   &  $18.90  \pm 0.03$\\  
                                                  & 0.17257 & 30    &  $19.03  \pm 0.03$\\ 
 & 0.18747 & 30     &  $19.16  \pm 0.03$	\\ 
 & 0.22093 & 30     &  $19.36  \pm 0.03$	\\  
 & 0.25426 & 30     &  $19.37  \pm 0.03$	\\  
 & 0.44461 & 30     &  $20.01  \pm 0.03$	\\  
 & 1.12910 & 60     &  $21.33  \pm 0.04$	\\  
 & 1.16363 & 60     &  $21.43  \pm 0.03$	\\ 
 & 1.19748 & 60     &  $21.46  \pm 0.03$	\\ 
 & 1.34318 & 60     &  $21.45  \pm 0.03$	\\  
 & 1.37520 & 60     &  $21.56  \pm 0.03$	\\ 
 & 1.43264 & 60     &  $21.71  \pm 0.03$	\\  
 & 2.36948 & 120    &  $22.40  \pm 0.05$\\ 
 \hline
 \xs, $R^{\rm (a)}$                      &        0.131725  & 20       & $18.91 \pm 0.02$ \\
 \hline
 ISAAC, $Ks$                    &  0.21572                        & 60                                & $18.70 \pm 0.10$ \\          
                                           & 0.27302                        &  60                               &  $19.01 \pm 0.09$ \\  
 \hline
 Gemini GMOS, $r$       &    6.26922                     & 900                   &    $ 23.25 \pm 0.04 $                 \\
                                          &  21.28034                     & 900                   &   $ 23.42 \pm 0.05  $                 \\
                                          &  23.39284                     & 720                   &     $ 23.40 \pm 0.10  $               \\
 \hline
FT-S, Bessel R$^{\rm (a)}$ & 0.66238      & 1800   & $20.82 \pm 0.02$ \\
                                                 & 0.75856      & 1800   & $20.99 \pm 0.03$ \\
                                                 & 0.82398     &  1800  & $21.23 \pm 0.03$ \\ 
                                                 & 3.71735     & 3600 & $23.03 \pm 0.09$ \\
 \hline
FT-S, Sloan i                         & 0.63924     & 1800   & $20.49 \pm 0.02$  \\ 
                                                & 0.73542     & 1800   & $20.71 \pm 0.03$ \\
                                                & 0.80083     & 1800   & $20.79 \pm 0.05$ \\
                                                & 0.88065     & 1800   & $20.96 \pm 0.04$ \\
                                                & 3.77179     &  900    & $22.38 \pm 0.42$ \\
                                                & 3.78300     &  900   & $22.42 \pm 0.35$ \\                                             
\hline
\end{tabular}
\normalsize
\end{table}
\setcounter{table}{2}
\begin{table}
 \centering
 \scriptsize
  \caption{ Continued.}
  \begin{tabular}{@{}llll@{}}
  \hline
UVOT white & 0.0017	& 149.8	  &  $15.33	\pm0.02$\\
& 0.0066        	& 19.8	  &  $16.70	\pm0.05$\\
& 0.0086		& 19.7	  &  $16.97	\pm0.05$\\
& 0.0108		& 149.7	  &  $17.24	\pm0.03$\\
& 0.0136		& 19.8	  &  $17.45	\pm0.06$\\
& 0.0693		& 199.8	  &  $19.07	\pm0.04$\\
& 0.1471		& 361	  &  $19.75	\pm0.05$\\
& 0.3406		& 1614.2  &  $20.59	\pm0.07$\\
& 1.0156		& 566.6	  &  $21.73	\pm0.18$\\
& 1.4510		& 40789.7 &  $22.66	\pm0.27$\\
& 1.9175		& 29362	  &  $22.86	\pm0.19$\\
& 2.5761		& 29994.8 &  $23.35	\pm0.28$\\
& 3.2865		& 46836	  &  $22.82	\pm0.18$\\
&3.9556			&58829.1  &  $22.64	\pm0.25$\\
&5.2311			&138900.4 &  $23.79	\pm0.36$\\
&6.6682			&98684.1  &  $23.26	\pm0.23$\\
&8.4084			&121624.7 & $>23.81$\\
\hline
UVOT $v$ 
&0.0072&	19.8	&     $16.34	\pm0.14$\\
&0.0092&	19.8	&     $16.84	\pm0.18$\\
&0.0132&	194.1   &     $16.92	\pm0.14$\\
&0.0740&	199.8   &     $18.51	\pm0.17$\\
&0.2803&	476	&     $19.45	\pm0.23$\\
&0.7414&	36383.3 &     $20.63	\pm0.26$\\
\hline
UVOT $b$ &0.0063&19.7   &$16.20 \pm0.08$\\
&0.0083&	19.7	&$16.45 \pm0.09$\\
&0.0143&	191.8	&$17.22 \pm0.09$\\
&0.1304&	1911.2	&$19.38 \pm0.14$\\
&0.1432&	299.8	&$19.19 \pm0.12$\\
&0.3259&	906.9	&$20.15 \pm0.15$\\
&0.4128&	686.9	&$20.15 \pm0.17$\\
&0.6129&	790.9	&$20.27 \pm0.18$\\
&0.9072&	16951.2	&$21.24 \pm0.34$\\
&1.0088&	603.3	&$20.75 \pm0.3$\\
\hline
UVOT $u$&0.0047&249.7	&$16.05\pm	0.03$\\
&0.0080&	19.7	&$16.71\pm	0.08$\\
&0.0140&	193.5	&$17.27\pm	0.08$\\
&0.0811&	199.7	&$19.00\pm	0.08$\\
&0.2119&	717.9	&$19.89\pm	0.08$\\
&0.4035&	907	&$20.57\pm	0.11$\\
&0.4797&	693	&$20.56\pm	0.13$\\
&0.6411&	7484.4	&$21.24\pm	0.14$\\
&0.8418&	7481.9	&$22.23\pm	0.35$\\
&1.5784&	63376.6	&$22.69\pm	0.32$\\
&2.2041&	34276.6	&$22.35\pm	0.35$\\
&2.9722&	87931	&$22.96\pm	0.3  $\\
&3.6073&	12000.2	&$22.40\pm	0.32$\\
&5.0600&	132684.3&$23.01\pm	0.35$\\
&7.4991&	278015.6&$23.33\pm	0.28$\\
\hline
UVOT $uvw1$&0.0077&19.8   	&$17.41 \pm0.13$\\
&0.0098&	   19.8	        &$17.70 \pm0.15$\\
&0.0137&	   193.7	&$17.83 \pm0.11$\\
&0.0788&	   199.7	&$19.80 \pm0.13$\\
&0.1885&	   434.9	&$20.27 \pm0.12$\\
&0.3930&	   899.7	&$20.96 \pm0.13$\\
&0.4704&	   899.8	&$21.35 \pm0.17$\\
&0.6312&	   7587.9	&$22.01 \pm0.19$\\
&0.9308&	  24671.6	&$22.34 \pm0.24$\\
\hline
UVOT $uvm2$&0.0095&19.8	&$17.66 \pm0.17$\\
&0.0124	&19.7	        &$18.05 \pm0.21$\\
&0.0764	&199.8	        &$20.02 \pm0.17$\\
&0.4599	&899.8	        &$21.29 \pm0.17$\\
&0.5460	&790.2	        &$21.70 \pm0.23$\\
&0.7027	&8389.3	        &$22.10 \pm0.21$\\
&0.8646	&899.7	        &$22.04 \pm0.28$\\
\hline
UVOT $uvw2$&0.0069	 &19.8	&$17.88	\pm0.15$\\
&0.0089	&19.8	        &$ 18.05\pm0.16$\\
&0.0139	&19.8	        &$ 18.52\pm0.20$\\
&0.0717	&199.7	        &$ 20.37\pm0.16$\\
&0.2581	&748.1	        &$ 21.27\pm0.14$\\
&0.5256	&899.8	        &$ 21.81\pm0.18$\\
&0.8314	&19056.1	&$ 23.05\pm0.32$\\
&1.8138	&102925.1      &$>21.90$\\
&3.5897	&191453.2      &$>22.51$\\
&7.1639	&325158.9      &$>22.50$\\
\hline
 \end{tabular}
\normalsize
\end{table}

\label{lastpage}

\end{document}